\documentclass[12pt]{article}

\usepackage{amssymb,amsfonts,amsthm}
\usepackage{amsmath}
\usepackage{bm}             
\usepackage[mathscr]{eucal}
\usepackage{cancel}
\usepackage{wasysym}

\def\be{\begin{equation}}
\def\ee{\end{equation}}
\def\bea{\begin{eqnarray}}
\def\eea{\end{eqnarray}}

\def\bna{\mbox{\boldmath $\nabla$}}
\def\brho{\mbox{\boldmath $\rho$}}

\def\la{\langle}
\def\ra{\rangle}
\def\hsp5{\hspace{5mm}}

\theoremstyle{remark}

\newcommand{\cA}{\mathcal{A}}
\newcommand{\cC}{\mathcal{C}}
\newcommand{\cH}{\mathcal{H}}

\newcommand{\sfrac}[2]{{\textstyle{#1\over#2}}}

\title{\sc A simplified structure for the second order cosmological perturbation equations}

\begin{document}

\author{ \\
{\Large\sc Claes Uggla}\thanks{Electronic address:
{\tt claes.uggla@kau.se}} \\[1ex]
Department of Physics, \\
University of Karlstad, S-651 88 Karlstad, Sweden
\and \\
{\Large\sc John Wainwright}\thanks{Electronic address:
{\tt jwainwri@uwaterloo.ca}} \\[1ex]
Department of Applied Mathematics, \\
University of Waterloo,Waterloo, ON, N2L 3G1, Canada \\[2ex] }

\maketitle

\begin{abstract}

Increasingly accurate observations of the cosmic microwave background and the
large scale distribution of galaxies necessitate the study of nonlinear
perturbations of Friedmann-Lemaitre cosmologies, whose equations are
notoriously complicated. In this paper we present a new derivation of the
governing equations for second order perturbations  within the framework of
the metric-based approach that is minimal, as regards amount of calculation
and length of expressions, and flexible, as regards choice of gauge and
stress-energy tensor. Because of their generality and the simplicity of their
structure our equations provide a convenient starting point for determining
the behaviour of nonlinear perturbations of FL cosmologies with any given
stress-energy content, using either the Poisson gauge or the uniform
curvature gauge.

\end{abstract}

\centerline{\bigskip\noindent PACS numbers: 04.20.-q, 98.80.-k,
98.80.Bp, 98.80.Jk}

\section{Introduction}

Over the past ten years there have been a number of significant developments
in the theory of nonlinear cosmological perturbations and its applications
that have transformed the area from a quiet backwater into one that is at the
forefront of research in cosmology. The main impetus for this resurgence has
been the availability  of increasingly accurate observations of the cosmic
microwave background (CMB) and of the large scale distributions of galaxies.
These observations necessitate the study of possible deviations from
linearity, for example, non-Gaussianity in the CMB anisotropies, using
nonlinear perturbations of FL cosmologies\footnote {We follow the
nomenclature of Wainwright and Ellis (1997):~a Friedmann-Lemaitre (FL)
cosmology is a Robertson-Walker (RW) geometry that satisfies Einstein's field
equations.} (see for example, Bartolo {\it et al} (2004a), Bartolo {\it et
al} (2010b), and Pitrou {\it et al} (2010)).

In order to motivate our work we give a brief overview of recent
developments. First, Bartolo and collaborators have written a series of
papers that apply second order perturbation theory to physical problems
relating to the early universe and to the CMB. Bartolo and collaborators
restrict their considerations \emph{ab initio} to purely scalar metric
perturbations at linear order and to a flat FL  background. Their main
theoretical tool is the set of expressions for the perturbed Einstein tensor
at second order for this class of perturbed metrics, derived without
restricting the gauge. They then introduce the Poisson gauge via
gauge-fixing. We refer to Acquaviva {\it et al} (2003) (equation (4) for the
perturbed metric and Appendix A5 for the perturbed Einstein
tensor.\footnote{See also Bartolo {\it et al} (2004a), equations (104) and
(A.36--(A.43).}). As regards physical applications they consider the case of
a scalar field (Acquaviva {\it et al} (2003)), a perfect fluid with linear
equation of state (Bartolo {\it et al} (2004b)), and dust and a cosmological
constant (Bartolo {\it et al} (2010a)). They use the expressions for the
perturbed Einstein tensor referred to above to obtain the governing equations
for the second order perturbations in these cases.

Second,  Malik and collaborators (see, Malik (2007), Malik {\it et al} (2008)
and Huston and Malik (2011)) have written a series of papers that apply
second order perturbation theory to FL cosmologies with one or more scalar
fields. They make the same simplifying assumptions as Bartolo and
collaborators, but instead use the uniform curvature gauge. They formulate
the governing equations in a way that is suitable for numerical computation.
Third, Nakamura (2003) introduced a geometrical method for constructing gauge
invariants for linear and nonlinear (second order) perturbations which he
later applied to derive the governing equations (see Nakamura (2006) and
Nakamura (2007)) effectively using the Poisson gauge. Finally, Noh and Hwang
(2004)  have given a comprehensive  treatment of second order perturbations,
with arbitrary spatial curvature and arbitrary stress-energy tensor,
including a scalar field and a perfect fluid as special cases. They use the
$3+1$-formulation of the Einstein equations and write the governing equations
in a so-called gauge-ready form.\footnote {We refer to Hwang and Noh (2007)
for further details and to Hwang {\it et al} (2012) for an application of the
formalism to determining second order perturbations of dust cosmologies.}

Despite the impressive progress that has been made to date the
theory of nonlinear cosmological perturbations nevertheless presents
challenges.  Indeed, the theory is notorious for tedious
calculations leading to lengthy quadratic expressions, the so-called {\it
source terms}, that can obscure the overall structure of the equations.
There is thus a need for a formulation of the governing equations that is
both general and concise and that will hence provide a suitable
starting point for future investigations.

Motivated by this state of affairs we present a new approach to the derivation of the governing equations
for second order perturbations  within the framework of the \emph{metric-based
approach}.\footnote{By this we mean the standard approach to cosmological
perturbations in which one  formulates the governing equations in terms of
gauge-invariant variables associated with the perturbed metric tensor and the
perturbed stress-energy tensor, using local coordinates.} We have designed
our approach to be \emph{minimal} in the sense that we calculate the least
number of objects and keep the number of lengthy expressions to a minimum,
thereby revealing useful mathematical structure. We
rely extensively on the notation and formalism for describing linear
perturbations introduced in two recent papers (Uggla and Wainwright (2011)
and (2012), hereafter referred to as UW1 and UW2, respectively). In UW1 we
gave the linearized Einstein equations with an arbitrary stress-energy tensor
in two different but complementary gauge-invariant forms, which we referred
to as the \emph{Poisson form}, associated with the work of Bardeen (1980),
and the \emph{uniform curvature form}, associated with the work of Kodama and
Sasaki (1984). In the present paper we derive and cast the perturbed Einstein
equations at second order into a convenient form analogous to the linearized equations
in UW1 but differing by the addition of source terms.

We begin our derivation by introducing so-called \emph{geometric perturbation
operators}, which provide the first level of structure by decomposing the
perturbed Riemann and Einstein tensors into a linear \emph{leading order
term} and a \emph{quadratic source term}, the latter being present only at
second order. A second level of structure is provided by the use of certain
linear combinations of the perturbed Einstein tensor components and their
derivatives, leading directly to a simple minimal set of governing equations
that is convenient for analysis. The use of a shorthand notation for
differential operators that occur frequently and reveal key mathematical structure
makes the equations much more
tractable. A third level of structure is provided by our strategy of
decomposing the general source terms into simpler pieces and identifying
common expressions, without expanding them fully. Special cases of the source
terms that appear in the literature can easily be extracted in a convenient
form from our general expressions.

The outline of the paper is as follows. In Section~\ref{sec:geomop} we
present `geometric perturbation operators' for the Riemann and Einstein
tensors up to second order, and specialize the key operators to the Poisson
gauge and the uniform curvature gauge. Section~\ref{sec:ein} gives the
perturbed Einstein field equations to second order in both the Poisson and
uniform curvature forms. We conclude the main part of the paper with a
discussion in Section~\ref{sec:disc}. Finally Appendix~\ref{app:der} contains
the definitions and equations needed to provide the background for the
results in the main part of the paper, as well as the so-called Replacement
Principle to second order, while Appendix~\ref{app:ucg} contains the detailed
expressions for the Einstein source terms assuming a purely scalar perturbation
at linear order, using the uniform curvature gauge.

\section{Geometric perturbation operators}\label{sec:geomop}

\subsection{Background}

Following standard cosmological perturbation theory, we consider a
1-parameter family of spacetimes $g_{ab}(\epsilon)$, where $g_{ab}(0)$, the
unperturbed metric, is a RW metric, and $\epsilon$ is referred to as the {\it
perturbation parameter}.\footnote{We use Latin letters $a,b,\dots,f$ to
denote abstract spacetime indices.} We assign physical dimension $length$ to
the scale factor $a$ of the RW metric and $(length)^2$ to $g_{ab}(\epsilon)$.
Then the conformal transformation
\be\label{bar_g}  g_{ab}(\epsilon) = a^2{\bar g}_{ab}(\epsilon),\ee
yields a dimensionless metric ${\bar g}_{ab}(\epsilon)$.

The Riemann tensor associated with the metric $g_{ab}(\epsilon)$ is a
function of $\epsilon$, denoted $R^{ab}\!_{cd}(\epsilon)$, as is the Einstein
tensor, $G^a\!_b(\epsilon)$. The stress-energy tensor of the matter
distribution is also assumed to be a function of $\epsilon$, denoted
$T^a\!_b(\epsilon)$. We include all these possibilities by considering a
1-parameter family of tensor fields $A(\epsilon)$, which we assume can be
expanded in powers of $\epsilon$, {\it i.e.} as a Taylor series:
\begin{subequations}
\be\label{taylor} \mathrm{A}(\epsilon) = {}^{(0)}\!\mathrm{A} +
\epsilon\,{}^{(1)}\!\mathrm{A} + \sfrac{1}{2}\epsilon^2\,
{}^{(2)}\!\mathrm{A} + \dots\, .\ee
The coefficients are given by
\be\label{perturb}  {}^{(0)}\!\mathrm{A} = A(0),  \qquad
{}^{(1)}\!\mathrm{A}= \left.\frac{\partial
\mathrm{A}}{\partial\epsilon}\right|_{\epsilon=0}, \qquad
{}^{(2)}\!\mathrm{A}= \left.\frac{\partial^2
\mathrm{A}}{\partial\epsilon^2}\right|_{\epsilon=0},  \quad
\dots, \ee
\end{subequations}
where  ${}^{(0)}\!A$ is called the {\it unperturbed value},
${}^{(1)}\!\mathrm{A}$ is called  the {\it first order (linear)
perturbation} and ${}^{(2)}\!\mathrm{A}$ is called  the {\it
second order perturbation} of $A(\epsilon)$.

 In particular we assume that we can expand the
conformal metric ${\bar g}_{ab}(\epsilon)$ in~\eqref{bar_g} in powers of
$\epsilon$,
\begin{subequations}\label{metricperturb}
\begin{equation} {\bar g}_{ab}(\epsilon) =
{}^{(0)}\!{\bar g}_{ab} + \epsilon\,  {}^{(1)}\!{\bar g}_{ab} +
\sfrac12 \epsilon^2\,  {}^{(2)}\!{\bar g}_{ab} +\dots\,.
\end{equation}
We label the coefficients as
\be\label{gamma,f}  \gamma_{ab} := {}^{(0)}\!{\bar g}_{ab} = {\bar
g}_{ab}(0), \quad  {}^{(1)}\!f_{ab} := {}^{(1)}\!{\bar g}_{ab} =
\left.\frac{\partial{\bar g}_{ab}}{\partial\epsilon}\right|_{\epsilon=0},
\quad {}^{(2)}\!f_{ab} := {}^{(2)}\!{\bar g}_{ab} =
\left.\frac{\partial^2{\bar g}_{ab}}{\partial\epsilon^2}\right|_{\epsilon=0},
\ee
\end{subequations}
which is consistent with~\eqref{perturb}. We refer to ${}^{(1)}\!f_{ab}$ as
the \emph{first order metric perturbation} and ${}^{(2)}\!f_{ab}$ as the
\emph{second order metric perturbation}. To simplify the notation, we will
denote ${}^{(1)}\!f_{ab}$ by $f_{ab}$ when there is no risk of confusion. The
conformal background metric $\gamma_{ab}$ and its inverse $\gamma^{ab}$ will
play an important role in that they are used to lower and raise indices on
perturbed objects.

\subsection{Perturbation operators for the Riemann and Einstein tensors}

We next expand the Riemann and Einstein tensors in the form
\begin{subequations} \label{expand_geom}
\begin{align}
R^{ab}\!_{cd}(\epsilon) &= {}^{(0)}\!R^{ab}\!_{cd} +
\epsilon {}^{(1)}\!R^{ab}\!_{cd} + \sfrac12 \epsilon^2
\,{}^{(2)}\!R^{ab}\!_{cd} + \dots\, ,\\
G^a\!_b(\epsilon) &= {}^{(0)}\!G^a\!_b + \epsilon
{}^{(1)}\!G^a\!_b + \sfrac12 \epsilon^2 \,{}^{(2)}\!G^a\!_b +
\dots\, .
\end{align}
\end{subequations}
To describe the perturbations of the spacetime geometry up to
second order we introduce dimensionless \emph{leading order
linear operators} and \emph{quadratic source term operators}
for the Riemann tensor $R^{ab}\!_{cd}$ and the Einstein tensor
\be \label{ein} G^a\!_b = R^{ac}\!_{bc} - \sfrac12 \delta^a\!_b R^{cd}\!_{cd}.  \ee
We refer to these operators as the \emph{geometric perturbation
operators} and use the notation ${\mathsf R}^{ab}\!_{cd}(\bullet)$
and ${\mathsf G}^{a}\!_{b}(\bullet)$ for the leading order
operators, and
 ${\mathcal R}^{ab}\!_{cd}(\bullet,\bullet)$ and ${\mathcal G}^{a}\!_{b}(\bullet,\bullet)$
 for the source term operators. These operators
determine the dependence of the linear and quadratic terms in the Taylor
series~\eqref{expand_geom} on the perturbations of the metric, through
equations of the following form:\footnote{Here and elsewhere we use the
shorthand notation ${}^{(r)}\!f$ for ${}^{(r)}\!f_{ab},\,r=1,2.$}
\begin{subequations}
\begin{xalignat}{2}
a^2{}^{(1)}\!R^{ab}\!_{cd} &= \mathsf{R}^{ab}\!_{cd}({}^{(1)}\!f), &\quad
a^2{}^{(2)}\!R^{ab}\!_{cd} &=
\mathsf{R}^{ab}\!_{cd}({}^{(2)}\!f) +
{\mathcal{R}}^{ab}\!_{cd}({}^{(1)}\!f,{}^{(1)}\!f),  \label{rie_operator} \\
a^2{}^{(1)}\!G^{a}\!_{b} &= \mathsf{G}^{a}\!_{b}({}^{(1)}\!f),
 &\quad a^2{}^{(2)}G^{a}\!_{b}
&= \mathsf{G}^{a}\!_{b}({}^{(2)}\!f) +
{\mathcal{G}}^{a}\!_{b}({}^{(1)}\!f,{}^{(1)}\!f). \label{einstein_operator}
\end{xalignat}
\end{subequations}
It is important that there is only one leading order operator for each
tensor: \emph {the same operator acts on both ${}^{(1)}\!f$ and
${}^{(2)}\!f$.} In Appendix~\ref{app:curvperturb} we derive concise
expressions for $\mathsf{R}^{ab}\!_{cd}\!\left({}^{(r)}\!f\right), r=1,2,$
and ${\mathcal{R}}^{ab}\!_{cd}({}^{(1)}\!f,{}^{(1)}\!f)$ (see equations
\eqref{Rie_leading} and~\eqref{Rie_source}).

\subsubsection{Local coordinates and differential operators}\label{sec:coords}

To proceed further we need to work in a coordinate frame so that we can
calculate time and spatial components separately. We thus
introduce local coordinates\footnote{We use Greek letters to
denote spacetime coordinate indices on the few occasions that
they occur, and we use Latin letters $i,j,k,m$ to denote
spatial coordinate indices, which are lowered and raised using
$\gamma_{ij}$ and its inverse  $\gamma^{ij}$, respectively.}
$x^\mu=(\eta,x^i)$, with $\eta$ being the usual conformal time
coordinate\footnote{Since we assigned $a$ to have physical
dimension $length$, the conformal time $\eta$ and the
conformal spatial line-element $\gamma_{ij}dx^idx^j$ are
dimensionless. We choose the $x^i$ to be dimensionless,
which implies that the $\gamma_{ij}$ are also dimensionless.}
for the RW metric $g_{ab}(0)$, and such that the
unperturbed conformal metric $\gamma_{ab} := \bar{g}_{ab}(0)$
has components
\be \gamma_{00}=-1\, ,\qquad \gamma_{0i}=0\, ,\qquad
\gamma_{ij}\, ,\ee
where $\gamma_{ij}$ is the metric of a spatial geometry of constant
curvature. The function $a=a(\eta)$ is the background cosmological
scale-factor, which determines the \emph{dimensionless background Hubble
scalar} ${\cal H}$ according to
\be\label{hubble} {\cal H} = \frac{a'}{a} = aH, \ee
where $H$ is the true background Hubble scalar. Here and elsewhere in this
paper $'$ denotes the derivative with respect to $\eta$ of a background
function that depends only on $\eta$. As in UW1 we will use the geometric
background scalars ${\cal A}_G$ and ${\cal C}_G^2$ that can be defined in
terms of ${\mathcal H}$ by the following equations:
\begin{equation}\label{A_G}
{\cA}_G := 2(-\cH' + \cH^2 + K), \qquad
\cA_G' = -(1 + 3\cC_G^2)\cH\cA_G
\end{equation}
(see UW1, equation (42)).

In order to formulate the perturbation equations concisely
we have found it helpful to introduce a shorthand notation for
certain differential operators that occur frequently and
clarify the structure of the equations. The first
operator determines the evolution of scalar perturbations
when using the Poisson gauge and is defined by
\begin{subequations} \label{cal_L}
\be \label{L_s}  {\bf {\cal L}} :=
\partial_\eta^2 + 3\!\left(1 + {\cal C}_G^2\right)\!{\cal
H}\partial_\eta +  {\cal H}^2{\cal B} -(1 + 3\cC_G^2)K, \ee
where
\be {\cal B} :=
\frac{2{\cal H}^\prime}{{\cal H}^2} + 1 + 3{\cal C}_G^2, \ee
\end{subequations}
(see UW1, equation (56)\footnote{See the final paragraph of section 3.2 in
UW1 for references to the literature where related forms of this operator
appear.}). This operator has the important property that it can be written as
the product of two first order differential operators:
\begin{subequations} \label{op2}
\be \label{L_factor}  {\cal L}(\bullet) = {\cal H}{\cal L}_A{\cal
L}_B\left(\frac{\bullet}{{\cal H}}\right), \ee
where
\be {\cal L}_A := \partial_\eta + {\cal H}{\cal B}, \qquad
{\cal L}_B := \partial_\eta + 2{\cal H}, \ee
\end{subequations}
(see UW1, equation (55) and UW2, equation (39)). These operators play a
central role, in particular when using the uniform curvature gauge.

We will make extensive use of the second order
spatial differential operators defined by
\begin{equation} \label{2order_D}
{\bf D}^2 := \gamma^{ij}{\bf D}_i{\bf D}_j, \qquad {\bf
D}_{ij} := {\bf D}_{(i}{\bf D}_{j)} - \sfrac13 \gamma_{ij}{\bf D}^2,
\end{equation}
where ${\bf D}_i$ denotes covariant differentiation
with respect to the spatial metric $\gamma_{ij}$.
We will also use the following shorthand notation
\be \label{DA}  \left({\bf D}A\right)^2:=({\bf D}^k A)( {\bf D}_kA), \ee
where $A$ is a scalar field.

\subsubsection{Minimal representation of the perturbed Einstein tensor}

In deriving the governing equations for linear and second order perturbations
in section~\ref{sec:ein} we follow the approach of UW1 and consider four
linear combinations of the components of the perturbed Einstein tensor and
their derivatives, which we denote by
\be ({}^{(r)}\!\hat{G}_{ij},\, {}^{(r)}\!{G},\, {}^{(r)}\!{G}_i, \, {}^{(r)}\!{G}^0\!_i), \quad r=1,2, \ee
where\footnote{Note that ${}^{(r)}{G}_{ij}\equiv\gamma_{ki}{}^{(r)}\! {G}^k\!_{j}$
is not symmetric when $r=2$. We
thus compensate by symmetrizing in~\eqref{ein_minimal1}, so that
${}^{(2)}\!\hat{G}_{ij}$, as defined by that equation, is symmetric.}
\begin{subequations}\label{ein_minimal}
\begin{align}
{}^{(r)}\!\hat{G}_{ij} &:= \gamma_{k(i}{}^{(r)}\! {G}^k\!_{j)} -\sfrac13\gamma_{ij}{}^{(r)}\!{G}^k\!_k , \label{ein_minimal1}\\
{}^{(r)}\!{G} &:= {\cal C}_G^2{}^{(r)}\!{G}^0\!_0 +
\sfrac13{}^{(r)}\!{G}^k\!_k, \label{ein_minimal2}\\
{}^{(r)}\!{G}_i &:=  - {\bf D}_i{}^{(r)}\!{G}^0\!_0 -
3{\cal H}{}^{(r)}\!{G}^0\!_i, \label{ein_minimal3}
\end{align}
\end{subequations}
and ${\mathcal C}_G^2$ is defined by~\eqref{A_G}. The use of these
combinations leads to a convenient form of the governing equations. In
particular, in the linear case ($r=1$) when using the Poisson gauge the
expression \eqref{ein_minimal2} leads directly to the \emph{ evolution
equation for the Bardeen potential} in terms of the differential operator
${\mathcal L}$ in~\eqref{cal_L}, while~\eqref{ein_minimal3} leads to the
\emph{generalized Poisson equation} that determines the density perturbation
in terms of the Bardeen potential (see UW1 equations (54)). We will see that
this pattern is repeated in the nonlinear case.

In order to calculate the Einstein combinations~\eqref{ein_minimal} we
express them in terms of the perturbations of the Riemann tensor
using~\eqref{ein} and~\eqref{expand_geom}:\footnote{Here and elsewhere $\la
.. \ra$ stands for a symmetrized and trace-free spatial index pair of a
perturbed object with respect to the  background spatial metric
$\gamma_{ij}$, while objects with a symmetrized and trace-free spatial index
pair are given a hat over the kernel symbol. \label{hat_angle}}
\begin{subequations}\label{ein_min_rie}
\begin{align}
{}^{(r)}\!\hat{G}_{ij} & =
\gamma_{k\la i}{}^{(r)}\!{R}^{\alpha k}\!_{|\alpha| j\ra}, \label{ein_rie_1p}\\
{}^{(r)}\!{G} & = - \sfrac16(1+ 3{\cal C}_G^2){}^{(r)}\!{R}^{pq}\!_{pq} -
\sfrac23{}^{(r)}\!{R}^{0k}\!_{0k}, \label{ein_rie_2p}\\
{}^{(r)}\!{G}_i & = \sfrac12 {\bf D}_i{}^{(r)}\!{R}^{pq}\!_{pq} -
3{\cal H}{}^{(r)}\!{R}^{0k}\!_{ik}, \label{ein_rie_3p}\\
{}^{(r)}\!{G}^0\!_i &\: = {}^{(r)}\!{R}^{0k}\!_{ik}. \label{ein_rie_4p}
\end{align}
\end{subequations}

It follows from \eqref{einstein_operator} that the dependence of the Einstein
combinations \eqref{ein_minimal} on the metric perturbations is given by
equations of the form
\begin{subequations} \label{ein_operator}
\begin{xalignat}{2}
a^2{}^{(1)}\!\hat{G}_{ij} &=
\hat{\mathsf{G}}_{ij}\!\left({}^{(1)}\!f\right)\!, &\quad
a^2{}^{(2)}\!\hat{G}_{ij} &= \hat{\mathsf{G}}_{ij}\!\left({}^{(2)}\!f\right)
+
{\mathcal{G}}_{ij}(f,f), \\
a^2{}^{(1)}\!{G} &= \mathsf{G}\!\left({}^{(1)}\!f\right)\!, &\quad
 a^2{}^{(2)}\!{G} &= \mathsf{G}\!\left({}^{(2)}\!f\right) +
{\mathcal{G}}(f,f), \\
a^2{}^{(1)}\!{G}_i &= \mathsf{G}_i\!\left({}^{(1)}\!f\right)\!, &\quad
a^2{}^{(2)}\!{G}_i &= \mathsf{G}_i\!\left({}^{(2)}\!f\right) +
{\mathcal{G}}_i(f,f),  \\
a^2{}^{(1)}\!{G}^0\!_i &= \mathsf{G}^{0}\!_{i}\!\left({}^{(1)}\!f\right)\!,
&\quad a^2{}^{(2)}\!{G}^0\!_i &=
\mathsf{G}^{0}\!_{i}\!\left({}^{(2)}\!f\right) +
{\mathcal{G}}^{0}\!_{i}(f,f),
\end{xalignat}
\end{subequations}
where for brevity we have denoted ${}^{(1)}\!f$ by $f$ in the source terms.

\subsubsection{The leading term operators}

We can express the leading order operators in~\eqref{ein_operator} in terms
of ${\mathsf{R}}^{ab}\!_{cd}(f)$ using equations \eqref{ein_min_rie}:
\begin{subequations}  \label{lead_G}
\begin{align}
\hat{\mathsf G}_{ij}(f) & = \gamma_{k\la i}{\mathsf R}^{\alpha k}\!_{|\alpha| j\ra}(f), \label{lead_ein_1}\\
{\mathsf G}(f) &= - \sfrac16(1+ 3{\cal C}_G^2){\mathsf R}^{pq}\!_{pq}(f) -
\sfrac23{\mathsf R}^{0k}\!_{0k}(f), \label{lead_ein_2}\\
{\mathsf G}_i(f)  &= \sfrac12 {\bf D}_i{\mathsf R}^{pq}\!_{pq}(f) -
3{\cal H}{\mathsf R}^{0k}\!_{ik}(f), \label{lead_ein_3}\\
{\mathsf G}^0\!_i(f) &\: = {\mathsf R}^{0k}\!_{ik}(f), \label{lead_ein_4}
\end{align}
\end{subequations}
where $f$ denotes ${}^{(1)}\!f$ or ${}^{(2)}\!f$. On using
equations~\eqref{Rie_leading} for ${\mathsf{R}}^{ab}\!_{cd}(f)$, in
conjunction with~\eqref{deriv}, we obtain after some manipulation:\footnote
{We refer to footnote \ref{hat_angle} for the $\la .. \ra$ and hat notation,
and to equations~\eqref{cal_L}--\eqref{2order_D} for the definitions of the
differential operators.}
\begin{subequations}\label{ein_leading}
\begin{align}
\hat{\mathsf G}_{ij}(f) &= \sfrac12{\bf D}_{ij}\!\left(f_{00} - \sfrac13 {f}^k\!_k\right)
+ {\cal L}_B\hat{\mathsf{Y}}_{ij}(f)
+ {\bf D}^k\!_{\la i}\hat{f}_{j\ra k} - \sfrac16({\bf D}^2 + 3K)\hat{f}_{ij}, \\
\begin{split}
{\mathsf G}(f) &= -\sfrac13\!\left[({\cal L} - {\cal C}_G^2{\bf D}^2){f}^k\!_k  +
\left(3{\cal H}{\cal L}_A + {\bf D}^2 \right)(f_{00} - \sfrac13 {f}^k\!_k)\right]\\
& \quad - \sfrac16(1 + 3{\cal C}_G^2){\bf D}^k\!_m \hat{f}^m\!_k +
\sfrac23\left( {\cal L}_B +3 {\cal H}\,{\cal C}_G^2\right){\bf D}^k f_{k0}, \end{split}\\
{\mathsf G}_i(f) &= - {\bf D}_i\!\left[ \sfrac13({\bf D}^2 + 3K){f}^k\!_k -
\sfrac12{\bf D}^k\!_m \hat{f}^m\!_k  \right]
+ 3{\cal H}{\bf D}^k\, \hat{\mathsf{Y}}_{ik}(f) , \\
{\mathsf G}^0\!_i(f) &= \sfrac23{\bf D}_i\mathsf{Y}(f) - {\bf D}^k\, \hat{\mathsf{Y}}_{ik}(f),
\end{align}
\end{subequations}
where
\be \mathsf{Y}(f) := \sfrac{3}{2}{\cal H}f_{00}
+ \sfrac{1}{2}\partial_\eta f^k\!_k  - {\bf D}^kf_{k0}, \quad
\hat{\mathsf{Y}}_{ij}(f) := \sfrac{1}{2}\partial_\eta
\hat{f}_{ij} - {\bf D}_{\la i}f_{j\ra 0},  \ee
and $f$ denotes ${}^{(1)}\!f$ or ${}^{(2)}\!f$. These equations are the main
result of this subsection and play a central role in our derivation of the
perturbation equations.

\subsubsection{The source term operators}

Calculating the  Einstein source terms ${\mathcal{G}}(f,f)$
in~\eqref{ein_operator} presents a major challenge. Our strategy is to first
express them in terms of ${\mathcal{R}}^{ab}\!_{cd}(f,f)$ using
equations~\eqref{ein_min_rie}:
\begin{subequations}  \label{source_G}
\begin{align}
\hat{\mathcal G}_{ij}(f,f) & = \gamma_{k\la i}{\mathcal R}^{\alpha k}\!_{|\alpha| j\ra}(f,f), \label{ein_1}\\
{\mathcal G}(f,f) &= - \sfrac16(1+ 3{\cal C}_G^2){\mathcal R}^{pq}\!_{pq}(f,f) -
\sfrac23{\mathcal R}^{0k}\!_{0k}(f,f), \label{ein_2}\\
{\mathcal G}_i(f,f)  &= \sfrac12 {\bf D}_i{\mathcal R}^{pq}\!_{pq}(f,f) -
3{\cal H}{\mathcal R}^{0k}\!_{ik}(f,f), \label{ein_3}\\
{\mathcal G}^0\!_i(f,f) &\: = {\mathcal R}^{0k}\!_{ik}(f,f). \label{ein_4}
\end{align}
\end{subequations}
We then substitute the expression for ${\mathcal{R}}^{ab}\!_{cd}(f,f)$ given
by equations~\eqref{Rie_source} in the Appendix into~\eqref{source_G}. This
leads to the expressions~\eqref{Einstein_source1} for the Einstein source
terms, each as a sum of simpler terms that can be calculated separately. The
constituent terms are given by
equations~\eqref{Einstein_source2}--\eqref{Q_sf}.

\subsection{Gauge invariants for the Einstein tensor}

We associate gauge invariants with the linear and second order perturbations
of any tensor using a method pioneered by Nakamura\footnote {See for example
Nakamura (2007), section 2.3.}, which we modify to ensure that the gauge
invariants are dimensionless. The process of construction, which involves
introducing co-called compensating gauge fields denoted by ${}^{(r)}X,\,
r=1,2,$ is described briefly in Appendix~\ref{app:repl}, and is specified by
equation~\eqref{bold_A} for an arbitrary tensor and by~\eqref{bold_f} for the
conformal metric tensor. The gauge invariants associated with the
perturbations of the conformal metric tensor and the Einstein tensor by this
process, which we refer to as $X$-compensation, are denoted by ${}^{(r)}{\bf
f}_{ab}[X]$ and ${}^{(r)}{\bf G}^a\!_b[X]$, respectively.

In equation~\eqref{einstein_operator} we expressed the perturbations
${}^{(r)}G^a\!_b,\, r=1,2,$ of the Einstein tensor in terms of the
perturbations ${}^{(r)}\!f_{ab}$ of the metric tensor in gauge-variant form,
using the Einstein geometric operators $ \mathsf{G}^{a}\!_{b}$ and
${\mathcal{G}}^{a}\!_{b}$. By applying the Replacement Principle\footnote
{Make the replacements ${}^{(r)}\!f\rightarrow{}^{(r)}{\bf f}[X],\, r=1,2,$
in~\eqref{einstein_operator}, in analogy with~\eqref{Rie_RP}.} to
equation~\eqref{einstein_operator} we  obtain the following expressions for
the Einstein gauge invariants in terms of the metric gauge invariants:
\be \label{ein_gi}  {}^{(1)}\!{\bf G}^{a}\!_{b}[X] =
\mathsf{G}^{a}\!_{b}(^{(1)}{\bf f}),\qquad {}^{(2)}\!{\bf G}^{a}\!_{b}[X] =
\mathsf{G}^{a}\!_{b}(^{(2)}{\bf f}) + {\mathcal{G}}^{a}\!_{b}(^{(1)}{\bf
f},^{(1)}\!{\bf f} ),    \ee
where we use the shorthand notation ${}^{(r)}{\bf f}\equiv{}^{(r)}{\bf
f}_{ab}[X],\,r=1,2$.

In this section we derive explicit expressions for the gauge invariants on
the right side of~\eqref{ein_gi} for two specific choices of the gauge fields
$X$. The starting point is to consider the gauge invariants associated with
the first and second order metric perturbations.

\subsubsection{Gauge invariants for the metric tensor perturbations}

To construct dimensionless gauge invariants associated with the linear
perturbation $f_{ab}$ of the conformal metric tensor we define\footnote {See
equation~\eqref{bold_f1} in this paper and  equation (16) in UW1.}
\be \label{f_bold} {\bf f}_{ab}[X] :=
f_{ab} - a^{-2}\pounds_{{}^{(1)}\!X}\!\left(a^2\gamma_{ab}\right), \ee
where the gauge field ${}^{(1)}\!X^a$ has to be chosen appropriately. In
order to construct a metric gauge field one has to decompose $f_{ab}$ into
scalar, vector and tensor modes, which we label as follows  (see UW1,
equation (18)):
\begin{subequations}\label{f_decomp}
\begin{align}
f_{00} &= -2\varphi, \\
f_{0i} &= {\bf D}_i B + B_i,\\
f_{ij} &= -2\psi \gamma_{ij} + 2{\bf D}_i {\bf D}_j C + 2{\bf D}_{(i} C_{j)} + 2C_{ij},
\end{align}
where the vectors $B_i$ and $C_i$ and the tensor $C_{ij}$
satisfy:
\begin{equation}
{\bf D}^i B_i = 0, \qquad  {\bf D}^i C_i = 0, \qquad
C^i\!_i = 0, \qquad  {\bf D}^i C_{ij} = 0.
\end{equation}
\end{subequations}
We use equations~\eqref{f_decomp} as a model for doing a mode decomposition of
${\bf f}_{ab}[X]$, using an obvious notation.\footnote
{For example, $\varphi\rightarrow\Phi[X], B\rightarrow{\bf B}[X]$,
as in equation~\eqref{modefX}.  }

As shown in UW1 there are two ways to choose $X$ uniquely in terms
of $f_{ab}$, leading to the {\it Poisson gauge field} $X_{\mathrm p}$ and the
{\it uniform curvature gauge field} $X_{\mathrm c}$. The corresponding
expressions for ${\bf f}_{ab}[X]$ are as follows (see UW1, equations
(28)--(31)):
\begin{subequations}  \label{bf_Poisson}
\begin{equation}
{\bf f}_{00}[X_\mathrm{p}] = -2  \Phi, \qquad {\bf f}_{0
i}[X_\mathrm{p}] =  {\bf B}_i, \qquad {\bf f}_{ij}[X_\mathrm{p}] = -2\Psi \gamma_{ij}
+ 2{\bf C}_{ij},
\end{equation}
where
\be \Phi:=\Phi[X_{\mathrm p}], \qquad  \Psi:=\Psi[X_{\mathrm p}], \ee
\end{subequations}
and
\begin{subequations}  \label{bf_curv}
\be {\bf f}_{00}[ X_{\mathrm{c}}] = -2{\bf A} ,\qquad
{\bf f}_{0i}[ X_{\mathrm{c}}] = {\bf D}_i {\bf B} +  {\bf B}_i , \qquad {\bf
f}_{ij}[ X_{\mathrm{c}}] = 2{\bf C}_{ij} , \ee
where\footnote
{In UW1 we introduced the symbols $A$ and $B$ for these gauge invariants,
following the notation of Kodama and Sasaki (1984), equations (3.4) and (3.5).}
\be\label{curv_scalar} {\bf A} := \Phi[ X_{\mathrm{c}}],
\qquad {\bf B} := {\bf B}[ X_{\mathrm{c}}]. \ee
\end{subequations}
In both case the vector and tensor modes satisfy
\be {\bf D}^i {\bf B}_i = 0, \qquad {\bf C}^i\!_i = 0, \qquad {\bf
D}^i{\bf C}_{ij} =0. \ee

To construct gauge invariants associated with the second order perturbation
${}^{(2)}\!f_{ab}$ we introduce a second gauge field ${}^{(2)}\!X$ and
define\footnote {See equation~\eqref{bold_f} in
Appendix~\ref{app:gauge_field}.}
\begin{subequations}
\be {}^{(2)}\!{\bf f}_{ab}[X] := {}^{(2)}\!f_{ab} - a^{-2}\pounds_{{}^{(2)}\!X}\!\left(a^2\gamma_{ab}\right) +
{\mathcal F}_{ab}[X],  \label{f2_X}  \ee
where
\be {\mathcal F}_{ab}[X] := - a^{-2}\pounds_{{}^{(1)}\!X}\!\left(2a^2 f_{ab} -
\pounds_{{}^{(1)}\!X}\!\left(a^2\gamma_{ab}\right)\right)\!. \label{F2X1} \ee
\end{subequations}
The key point is that one can construct a gauge field ${}^{(2)}\!X_{\mathrm
p}$ such that  ${}^{(2)}\!{\bf f}_{ab}[X_{\mathrm p}]$ {\it has the same form
as} ${\bf f}_{ab}[X_{\mathrm p}]$ in~\eqref{bf_Poisson}.  In other words,
${}^{(2)}\!{\bf f}_{ab}[X_\mathrm{p}]$ can be obtained by making the
substitutions
\begin{equation}\label{bf2_Poisson} \Phi\rightarrow {}^{(2)}\Phi, \qquad  \Psi\rightarrow
{}^{(2)}\Psi, \qquad {\bf B}_i\rightarrow {}^{(2)}{\bf B}_i[X_\mathrm{p}], \qquad
{\bf C}_{ij}\rightarrow{}^{(2)}{\bf C}_{ij}[X_\mathrm{p}],
\end{equation}
in~\eqref{bf_Poisson}. Similarly, one can construct a gauge field
${}^{(2)}\!X_{\mathrm c}$ such that ${}^{(2)}\!{\bf f}_{ab}[X_{\mathrm c}]$
is obtained by making the substitutions
\begin{equation}\label{bf2_curv} {\bf A}\rightarrow {}^{(2)}{\bf A}, \qquad  {\bf B}\rightarrow
{}^{(2)}{\bf B}, \qquad {\bf B}_i\rightarrow {}^{(2)}{\bf B}_i[X_\mathrm{c}], \qquad
{\bf C}_{ij}\rightarrow{}^{(2)}{\bf C}_{ij}[X_\mathrm{c}],
\end{equation}
in~\eqref{bf_curv}. Details about the construction of these gauge fields and
the expressions for the metric gauge invariants in
equations~\eqref{bf_Poisson},~\eqref{bf_curv},~\eqref{bf2_Poisson}
and~\eqref{bf2_curv} in terms of the gauge-variant metric perturbations are
given in Appendix~\ref{app:gauge_field}. It is important to note, however,
that {\it these explicit expressions are not required in what follows.} All
that is required is the general form of ${\bf f}_{ab}[X]$ and ${}^{(2)}\!{\bf
f}_{ab}[X]$ in the Poisson gauge and in the uniform curvature gauge, as given
by equations~\eqref{bf_Poisson},~\eqref{bf_curv},~\eqref{bf2_Poisson}
and~\eqref{bf2_curv}.

\subsubsection{The leading order terms}

To obtain gauge-invariant expressions for the leading order terms we simply
make the substitutions $f_{ab} \rightarrow {\bf f}_{ab}[X],\,{}^{(2)}\!f_{ab}
\rightarrow {}^{(2)}\!{\bf f}_{ab}[X]$ in equations~\eqref{ein_leading}. For
the {\it Poisson gauge} we use~\eqref{bf_Poisson} which gives the leading
order Einstein operator ${\mathsf G}$ acting on the first order metric
perturbation ${\bf f}_\mathrm{p}\equiv{\bf f}_{ab}[X_\mathrm{p}]$. After some
manipulation we obtain:\footnote {The identities (B.39b) and (B.39f) in UW1
are needed.}
\begin{subequations}\label{leadein_Poisson}
\begin{align}
\hat{\mathsf{G}}_{ij}({\bf f}_\mathrm{p}) &=
{\bf D}_{ij}\!\left(\Psi - \Phi\right) -
{\bf D}_{(i}{\cal L}_B {\bf B}_{j)} +
\left({\cal L}_B\partial_\eta + 2K - {\bf D}^2\right){\bf C}_{ij},\\
\mathsf{G}({\bf f}_\mathrm{p}) &= 2\!\left[\left({\cal L} -
{\cal C}_G^2{\bf D}^2\right)\Psi
-\left({\cal H}{\cal L}_A + \sfrac13{\bf D}^2\right)\left(\Psi - \Phi\right)\right]\! ,\\
\mathsf{G}_i({\bf f}_\mathrm{p}) &= 2{\bf D}_i({\bf D}^2 + 3K)\Psi -
\sfrac32{\cal H}({\bf D}^2 + 2K){\bf B}_i ,\\
\mathsf{G}^0\!_i({\bf f}_\mathrm{p}) &=
 -2{\bf D}_i\!\left(\partial_\eta\Psi + {\cal H}\Phi\right)
+ \sfrac12({\bf D}^2 + 2K){\bf B}_i .
\end{align}
\end{subequations}
To obtain ${\mathsf G}$ acting on the second order metric perturbation
${}^{(2)}\!{\bf f}_\mathrm{p}\equiv\!{}^{(2)}\!{\bf f}_{ab}[X_\mathrm{p}]$ we
simply make the replacements~\eqref{bf2_Poisson} in~\eqref{leadein_Poisson}.

Similarly for the {\it uniform curvature gauge} we use~\eqref{bf_curv}, which
gives the leading order Einstein operator ${\mathsf G}$ acting on the first
order metric perturbation ${\bf f}_\mathrm{c}\equiv{\bf
f}_{ab}[X_\mathrm{c}]$. After some manipulation we obtain:\footnote{The terms
involving the vector and tensor modes are the same as in the Poisson case.}
\begin{subequations}\label{leadein_curv}
\begin{align}
\hat{\mathsf G}_{ij}({\bf f}_\mathrm{c}) &=- {\bf D}_{ij} ({\mathcal L}_B{\bf B} + {\bf A})
- {\bf D}_{(i} {\mathcal L}_B {\bf B}_{j)} +
\left({\mathcal L}_B\partial_\eta  + 2K - {\bf D}^2\right){\bf C}_{ij}, \\
{\mathsf G}({\bf f}_\mathrm{c}) &=   2{\cal H}[{\mathcal L}_A{\bf A} +
{\cal C}_G^2 {\bf D}^2{\bf B}] +\sfrac23{\bf D}^2 ({\mathcal L}_B{\bf B} + {\bf A}), \\
{\mathsf G}_i({\bf f}_\mathrm{c}) &= - 2 \cH{\bf D}_i({\bf D}^2 + 3K){\bf B} - \sfrac32\cH({\bf D}^2 + 2K){\bf B}_i, \\
\mathsf{G}^0\!_i({\bf f}_\mathrm{c}) &=
 -2{\bf D}_i\!\left( {\cal H}{\bf A} - K{\bf B}\right)
+ \sfrac12({\bf D}^2 + 2K){\bf B}_i .
\end{align}
\end{subequations}
To obtain ${\mathsf G}$ acting on the second order metric perturbation
${}^{(2)}\!{\bf f}_\mathrm{c}\equiv\!{}^{(2)}\!{\bf f}_{ab}[X_\mathrm{c}]$ we
simply make the replacements~\eqref{bf2_curv} in~\eqref{leadein_curv}.

Equations~\eqref{leadein_Poisson} and~\eqref{leadein_curv} provide the
leading order terms in the perturbed Einstein equations~\eqref{ein3} at
linear and second order using the Poisson gauge and the uniform curvature
gauge, respectively. For the reader's convenience we note that the various
differential operators are defined by
equations~\eqref{cal_L}--\eqref{2order_D}.

\subsubsection{The source terms}

The  source terms are given in general in gauge-variant form
by~\eqref{Einstein_source1} in conjunction
with~\eqref{Einstein_source2}--\eqref{Q_sf}. They are obtained in
gauge-invariant form by simply making the replacement $f_{ab} \rightarrow
{\bf f}_{ab}[X]$ in these equations, in particular with ${\bf f}_{ab}[X_{\mathrm p}]$
for the Poisson gauge and ${\bf f}_{ab}[X_{\mathrm c}]$ for the uniform curvature gauge
(see equations~\eqref{bf_Poisson} and~\eqref{bf_curv}).
 To illustrate
our approach we  consider a popular special case, namely the
Poisson gauge with the metric perturbation restricted as follows:

\vspace{3mm} \noindent \emph{Metric assumptions}: The  vector
and tensor modes of the metric perturbation are zero at first order, {\it i.e.}
\begin{subequations}\label{m_assump}
\begin{equation} \label{m_assump1}
{\bf B}_i=0, \qquad {\bf C}_{ij}=0,
\end{equation}
and in addition the scalar mode at first order satisfies
\begin{equation}
\Phi = \Psi.
\end{equation}
\end{subequations}
Subject to these assumptions \eqref{bf_Poisson} reduces to
\begin{equation} \label{Poissondef2} {\bf f}_{00}[X_{\mathrm p}]=-2\Psi, \qquad {\bf
f}_{0i}[X_{\mathrm p}]=0, \qquad {\bf f}_{ij}[X_{\mathrm p}]=
-2\Psi \gamma_{ij},
\end{equation}
where $\Psi$ is the {\it Bardeen potential}. We now make the replacement
$f_{ab} \rightarrow {\bf f}_{ab}[X_{\mathrm p}]$ in the
expressions~\eqref{Einstein_source1} for the Einstein source terms using the
special metric perturbation~\eqref{Poissondef2}. The constituent terms, as
given by equations~\eqref{Einstein_source2}--\eqref{Q_sf}, can be evaluated
separately and one finds that many terms are zero. This calculation yields
the following simple expressions for the source terms:
\begin{subequations}\label{special_source}
\begin{align}
\hat{\mathcal{G}}_{ij}({\bf f}_\mathrm{p},{\bf f}_\mathrm{p}) &=
4\!\left({\bf D}_{ij}\Psi^2 - ({\bf D}_{\la i}\Psi)({\bf D}_{j\ra}\Psi)\right),\\
\mathcal{G}({\bf f}_\mathrm{p},{\bf f}_\mathrm{p}) &=
- \sfrac13\!\left(1 + 3{\cal C}_G^2\right) {\mathcal R}(\Psi,\Psi) -
\sfrac83({\bf D}\Psi)^2 - 8{\cal H}{\cal L}_A\Psi^2 ,\\
\mathcal{G}_i({\bf f}_\mathrm{p},{\bf f}_\mathrm{p}) &=
{\bf D}_i  {\mathcal R}(\Psi,\Psi) + 12{\cal H}\!\left(\partial_\eta\Psi\right)\!{\bf D}_i\Psi , \\
\mathcal{G}^0\!_i({\bf f}_\mathrm{p},{\bf f}_\mathrm{p}) &=
 8{\cal H}{\bf D}_i\Psi^2 -4(\partial_\eta\Psi)({\bf D}_i\Psi),
\end{align}
where
\be \label{cal_R}
{\mathcal R}(\Psi,\Psi) := 2\!\left[ 3(\partial_\eta \Psi)^2  -
5({\bf D}\Psi)^2 + 4\!\left({\bf D}^2 + 3K\right)\!\Psi^2\right].
\ee
\end{subequations}
Note that the source terms are {\it quadratic expressions in the Bardeen
potential $\Psi$ and its derivatives}. Nakamura (2007) has given the source
terms in this case in the form ${\mathcal G}^a\!_b({\bf f}_\mathrm{p},{\bf
f}_\mathrm{p})$. We find complete agreement with his equations (6.13)--(6.16)
when they are transformed into the form~\eqref{special_source}.

In summary, equations~\eqref{leadein_Poisson} and~\eqref{special_source} give
the constituent geometrical parts in the second order field
equations~\eqref{ein3} in section~\ref{sec:ein}. They thus provide the
foundation for determining second order metric perturbations in the Poisson
gauge, subject to the simplifying metric assumption~\eqref{m_assump}.

Finally we note that one can similarly derive expressions for the source
terms using the uniform curvature gauge, subject to the simplifying
assumption~\eqref{m_assump1}. The resulting expressions are more complicated
than~\eqref{special_source} and so we give them in Appendix~\ref{app:ucg}.

\section{Perturbed Einstein equations}\label{sec:ein}

\subsection{General structure of the governing equations}

At zeroth order the non-zero components of Einstein's field
equations are given by\footnote{See, for example, Mukhanov {\it
et al} (1992), equation (4.2), noting the difference in
signature. We use units $c=1$ and $8\pi G=1$, where $c$ is the
speed of light and $G$ is the gravitational constant.}
\begin{subequations}\label{GT0}
\begin{align}
a^2\,{}^{(0)}\!G^0\!_0 &= -3({\cal H}^2 + K) \hskip-0.8cm
&=&\,\, - a^2{}^{(0)}\!\rho\, \hskip-3.2cm &=&\,\,\, a^2\,{}^{(0)}\!T^0\!_0,\\
a^2\, {}^{(0)}\!G^i\!_j &= -(2{\cal H}^\prime + {\cal H}^2 +
K)\delta^i\!_j \hskip-2.0cm &=& \,\,\, a^2{}^{(0)}\!p\,\delta^i\!_j\, \hskip-2cm
&=&\,\,\, a^2{}^{(0)}\!T^i\!_j,
\end{align}
\end{subequations}
where ${\cal H}$ is given by~\eqref{hubble} and $K$ is the curvature index,
defined in equation~\eqref{threecurv} in Appendix~\ref{app:exactcurv}.

The perturbed Einstein equations at linear and second order are
given by
\begin{equation}
{}^{(1)}G^a\!_b = {}^{(1)}T^a\!_b, \qquad {}^{(2)}G^a\!_b
= {}^{(2)}T^a\!_b.
\end{equation}
Assuming that the background Einstein equations are satisfied we
can write these equations in terms of the gauge invariants
${}^{(r)}{\bf G}^a\!_b[X]$ and ${}^{(r)}{\bf T}^a\!_b[X]$:
\be {}^{(1)}{\bf G}^a\!_b[X]={}^{(1)}{\bf T}^a\!_b[X], \qquad {}^{(2)}{\bf
G}^a\!_b[X]={}^{(2)}{\bf T}^a\!_b[X], \ee
as follows from the definition~\eqref{bold_A} in Appendix~\ref{app:repl}.
We can now use~\eqref{ein_gi} to express the left side of these
equations in terms of the Einstein operators:
\begin{subequations} \label{ein2}
\begin{gather}
{\mathsf G}^a\!_b({}^{(1)}{\bf f}) = {}^{(1)}{\bf
T}^a\!_b[X],\\
\mathsf{G}^{a}\!_{b}\!\left({}^{(2)}{\bf f}\right) +
{\mathcal{G}}^{a}\!_{b}({}^{(1)}{\bf f},\!{}^{(1)}{\bf f}) = {}^{(2)}{\bf
T}^a\!_b[X].
\end{gather}
\end{subequations}
At this stage we are  considering an arbitrary stress-energy tensor, whose
components are regarded as primary objects, {\it i.e.} they are not
constructed from other quantities as in the case of the Einstein tensor.
Before continuing we note that equations~\eqref{ein2} correspond to equations
(2.49) and (2.50) in Nakamura (2007).\footnote {See also, equations (38) and
(39) in Nakamura (2006). Nakamura's metric gauge invariants are related to
ours according to ${\mathcal L}_{ab} =a^2\,{}^{(1)}{\bf f}_{ab}[X],
\,{\mathcal H}_{ab} =a^2\,{}^{(2)}{\bf f}_{ab}[X]$.}

We next consider the combinations of equations~\eqref{ein2}
corresponding to the combinations of the Einstein components
defined in~\eqref{ein_minimal}:
\begin{subequations}  \label{ein3}
\begin{xalignat}{2}
{\hat{\mathsf G}}_{ij} ({}^{(1)} {\bf f}) &=
{}^{(1)}{\hat{\bf T}}_{ij} [X], &\quad
{\hat{\mathsf G}}_{ij}  ({}^{(2)} {\bf f}) +
{\hat{\mathcal G}} _{ij} ({\bf f}, {\bf f}) &= {}^{(2)}{\hat{\bf T}}_{ij} [X],\\
{\mathsf G} ({}^{(1)} {\bf f}) &=  {}^{(1)}{\bf T}[X], &\quad
{\mathsf G} ({}^{(2)} {\bf f}) + {\mathcal G} ({\bf f}, {\bf f}) &= {}^{(2)}{\bf T}[X],\\
{\mathsf G}_i ({}^{(1)} {\bf f}) &=  {}^{(1)}{\bf T}_i[X],
&\quad
{\mathsf G}_i ({}^{(2)} {\bf f}) + {\mathcal G} ({\bf f}, {\bf f}) &= {}^{(2)}{\bf T}_i[X],\\
{\mathsf G}^0\!_i ({}^{(1)}{\bf f}) &=  {}^{(1)}{\bf
T}^0\!_i[X], &\quad {\mathsf G}^0\!_i ({}^{(2)} {\bf f}) +
{\mathcal G}^0\!_i ({\bf f}, {\bf f}) &= {}^{(2)}{\bf
T}^0\!_i[X].
\end{xalignat}
\end{subequations}
The linear combinations of ${}^{(r)}{\bf T}^a\!_b[X]$ in~\eqref{ein3} are defined
in analogy with~\eqref{ein_minimal}
by\footnote{At first order ${}^{(1)}\!\hat{\bf T}_{ij}[X]$,
${}^{(1)}\!{\bf T}[X]$, ${}^{(1)}{\bf T}_i[X]$ are
intrinsic gauge invariants since they do not depend on the
choice of gauge field and hence can be written as
${}^{(1)}\!\hat{\bf T}_{ij}$, ${}^{(1)}\!{\bf T}$, ${}^{(1)}{\bf T}_i$
(see UW1, section 2.3). At second order this is no longer the case.}
\begin{subequations}
\begin{align}
{}^{(r)}\!\hat{\bf T}_{ij}[X] &:=
\gamma_{k(i}{}^{(r)}\!{\bf T}^k\!_{j)}[X] -
\sfrac13\gamma_{ij}{}^{(r)}\!{\bf T}^k\!_k[X],\\
{}^{(r)}\!{\bf T}[X] &: =
{\cal C}_T^2{}^{(r)}\!{\bf T}^0\!_0[X] + \sfrac13{}^{(r)}\!{\bf T}^k\!_k[X], \label{gamma}\\
{}^{(r)}\!{\bf T}_i[X] &:=
- {\bf D}_i{}^{(r)}\!{\bf T}^0\!_0[X] - 3{\cal H}{}^{(r)}\!{\bf T}^0\!_i[X], \label{T_i}
\end{align}
\end{subequations}
where
\begin{equation}
{\cal C}_T^2 = \frac{{}^{(0)}\!p^\prime}{{}^{(0)}\!\rho^\prime},
\qquad   {}^{(0)}\!p = \sfrac13 {}^{(0)}T^k\!_k,\qquad
{}^{(0)}\!\rho = -{}^{(0)}\!T^0\!_0 .
\end{equation}

Equations~\eqref{ein3} give a convenient \emph{minimal form of the governing
equations for linear and second order perturbations for any choice of gauge
field $X$ and any stress-energy tensor.} The leading order terms ${\mathsf
G}({\bf f})$, where ${\bf f}={}^{(1)}{\bf f}$ or ${}^{(2)}{\bf f}$, are given
by~\eqref{leadein_Poisson} for the Poisson gauge $X=X_{\mathrm p}$ and
by~\eqref{leadein_curv} for the uniform curvature  gauge $X=X_{\mathrm c}$.
The source terms ${\mathcal G}({\bf f}, {\bf f})$ are obtained in general by
making the substitution $f_{ab} \rightarrow {\bf f}_{ab}[X]$ in
equations~\eqref{Einstein_source1}--\eqref{Q_sf} for an arbitrary $X$, and
are given directly by~\eqref{special_source} for the special metric
perturbation~\eqref{m_assump} when using the Poisson gauge.

The final step is to decompose the governing equations~\eqref{ein3} into
equations for the scalar mode, the vector mode and the tensor mode. As with
the metric we perform a mode decomposition of the stress-energy gauge
invariants:\footnote {For brevity we drop the argument $[X]$ for the various
mode terms.}
\begin{subequations}\label{Tmode}
\begin{align}
{}^{(r)}\!\hat{\bf T}_{ij}[X] &= {\bf D}_{ij} {}^{(r)}\!\Pi +
2{\bf D}_{(i}{}^{(r)}\!\Pi_{j)} + {}^{(r)}\!{\Pi}_{ij} , \label{Thatmode}\\
{}^{(r)}\!{\bf T}_i[X] & = {\bf D}_i {}^{(r)}\!\Delta + {}^{(r)}\!\Delta_i, \label{Deltamode} \\
{}^{(r)}\!{\bf T}^0\!_i[X] & = {\bf D}_i {}^{(r)}\!V + {}^{(r)}\!\tilde{V}_i, \label{Vmode} \\
{}^{(r)}\!{\bf T}[X] & = {}^{(r)}\!\Gamma, \label{Gamma_mode}
\end{align}
\end{subequations}
where
\begin{equation}
{\bf D}^i{}^{(r)}\!\Pi_i =0 ,\quad {}^{(r)}\!{\Pi}^k\!_k = 0 ,\quad  {\bf
D}_i{}^{(r)}\!{\Pi}^i\!_j = 0, \quad {\bf D}^i {}^{(r)}\!\Delta_i = 0, \quad
{\bf D}^i {}^{(r)}\!\tilde{V}_i=0.
\end{equation}
A difficulty arises that is not present at the linear level. The leading
order terms ${\mathsf G}({}^{(2)}{\bf f}_{\mathrm p})$ and the stress-energy
terms ${}^{(2)}{\bf T}[X]$ in~\eqref{ein3} are expressed explicitly as a sum
of a scalar term, a vector term and a tensor term. On the other hand, {\it
the source terms ${\mathcal G}({\bf f},{\bf f})$ do not have this form}, as
can be seen, for example, from~\eqref{special_source}. One thus has to apply
what we call \emph{mode extraction operators} to~\eqref{ein3} in order to
separate the modes in the source terms. These operators are defined
in~\eqref{modeextractop}, using the letters ${\mathcal S}, {\mathcal V}$ and
${\mathcal T}$ to denote the scalar, vector and tensor modes, respectively.

For later use we note that applying the mode extraction operators
to~\eqref{Tmode} yields
\begin{subequations} \label{Tdecomp}
\begin{xalignat}{3}
{}^{(r)}\!\Pi &= {\cal S}^{ij} {}^{(r)}\!\hat{\bf T}_{ij}[X],
&\quad {}^{(r)}\!\Pi_i &= {\cal V}_{i}\!^{jk}{}^{(r)}\!\hat{\bf
T}_{jk}[X], &\quad {}^{(r)}\!\Pi_{ij} &= {\cal
T}_{ij}\!^{pq}{}^{(r)}\!\hat{\bf T}_{pq}[X],  \label{Tdecomp1} \\
{}^{(r)}\!\Delta &= {\cal S}^{i}{}^{(r)}\!{\bf T}_i[X] ,&\quad
{}^{(r)}\!\Delta_i &= {\cal V}_i\!^{j}
{}^{(r)}\!{\bf T}_j[X], && \\
{}^{(r)}\!V &= {\cal S}^{i}{}^{(r)}\!{\bf T}^0\!_i[X] ,&\quad
{}^{(r)}\!\tilde{V}_i &= {\cal V}_i\!^{j} {}^{(r)}\!{\bf
T}^0\!_j[X]. &&
\end{xalignat}
\end{subequations}
%

\subsection{The mode-decomposed governing equations}

In this section we give the mode-decomposed form of the governing Einstein
field equations at second order for perturbations of an FL cosmology with
arbitrary matter content, first using Poisson gauge invariants, and then
using uniform curvature gauge invariants. The source terms, identified by the
kernel ${\mathcal G}$, are obtained by making the substitution
$f\rightarrow{\bf f}_{\mathrm p}$ or $f\rightarrow{\bf f}_{\mathrm c}$ in
equations~\eqref{Einstein_source1}--\eqref{Q_sf}, or directly
by~\eqref{special_source} for the special metric
perturbation~\eqref{m_assump} when using the Poisson gauge. We note that, in
accordance with~\eqref{ein3}, {\it the governing equations at first order can
be simply obtained from the equations at second order by dropping the source
terms}, indicated by the kernel ${\mathcal G}$, and dropping the exponent
${}^{(2)}$ (or replacing it by ${}^{(1)}$).

\subsubsection*{The Poisson form}

To obtain the Poisson form we substitute~\eqref{leadein_Poisson}
and~\eqref{Tmode} into~\eqref{ein3} and then apply the mode extraction
operators~\eqref{modeextractop}, which leads to: \vspace{5mm}

\emph{Scalar mode}
\begin{subequations} \label{gov_scalar}
\begin{align}
{}^{(2)}\Psi - {}^{(2)}\Phi &= {}^{(2)}\Pi_{\mathrm p} - {\cal
S}^{ij}\, \hat{\mathcal{G}}{}_{ij}({\bf f}_{\mathrm p},{\bf f}_{\mathrm p}),\label{gov_scalar1}  \\
\begin{split}   
\left({\bf {\cal L}} - {\cal C}_G^2{\bf D}^2\right){}^{(2)}\Psi
&= \sfrac12{}^{(2)}\Gamma_{\mathrm p} + \\
&\quad \left({\cal H} {\mathcal L}_A +\sfrac13{\bf D}^2\right)(  {}^{(2)}\Pi_{\mathrm p} - {\cal
S}^{ij}\, \hat{\mathcal{G}}{}_{ij}({\bf f}_{\mathrm p},{\bf f}_{\mathrm p}))  -
\sfrac12\mathcal{G}({\bf f}_{\mathrm p},{\bf f}_{\mathrm p})  ,   \label{psi_evol}
\end{split} \\
({\bf D}^2 + 3K){}^{(2)}\Psi &= \sfrac12 {}^{(2)}\Delta_{\mathrm p}  -  \sfrac12{\cal
S}^i\,\mathcal{G}_i({\bf f}_{\mathrm p},{\bf f}_{\mathrm p}), \label{matter}\\
\partial_{\eta}{}^{(2)}\Psi + {\cal H}{}^{(2)}\Phi &= - \sfrac12{}^{(2)}V_{\mathrm p}  + \sfrac12{\cal
S}^i\,\mathcal{G}^0\!_i({\bf f}_{\mathrm p},{\bf f}_{\mathrm p}), \label{gov_scalar4}
\end{align}
\end{subequations}

\emph{Vector mode}
\begin{subequations}\label{gov_vector}
\begin{align}
\mbox{\boldmath{${\cal L}$}}_B{}^{(2)}\!{\bf B}_i[X_{\mathrm p}] &= - 2{}^{(2)}\!\Pi_i[X_{\mathrm p}] +
2{\cal V}_i\!^{jk}\, \hat{\mathcal{G}}{}_{jk}({\bf f}_{\mathrm p},{\bf f}_{\mathrm p}) ,  \label{vector_evol}\\
({\bf D}^2 + 2K){}^{(2)}\!{\bf B}_i[X_{\mathrm p}] &=  2{}^{(2)}\!\tilde{V}_i[X_{\mathrm p}] -
2{\cal V}_i\!^j\,\mathcal{G}^0\!_j({\bf f}_{\mathrm p},{\bf f}_{\mathrm p}), \label{vector_V1}
\end{align}
\end{subequations}

\emph{Tensor mode}
\begin{equation}\label{gov_tensor}
\left({\cal L}_B\partial_\eta + 2K - {\bf D}^2\right)\!{}^{(2)}\!{\bf C}_{ij}[X_{\mathrm p}] =
{}^{(2)}\!{\Pi}_{ij}[X_{\mathrm p}] - {\cal T}_{ij}\!^{km}\,
\hat{\mathcal{G}}_{km}({\bf f}_{\mathrm p},{\bf f}_{\mathrm p}).
\end{equation}
Here ${}^{(2)}\Psi,{}^{(2)}\Phi, {}^{(2)}\!{\bf B}_i[X_{\mathrm p}]$ and
${}^{(2)}\!{\bf C}_{ij}[X_{\mathrm p}]$ are the second order Poisson metric
gauge invariants and ${\bf f}_{\mathrm p}$ is shorthand for the first order
metric perturbation ${\bf f}_{ab}[X_{\mathrm p}]$ as given by~\eqref{bf_Poisson}.

The evolution of the scalar perturbations is
governed by equation \eqref{psi_evol}, a second order
partial differential equation for $^{(2)}\Psi$.
 In order to obtain a solution one first has to solve the
linearized field equations for the first order gauge-invariant metric
perturbation ${\bf f}_{ab}[X_{\mathrm p}]$, which then determines the
Einstein source terms. Once the second order matter terms
${}^{(2)}\Gamma_{\mathrm p}$ and ${}^{(2)}\Pi_{\mathrm p} $ have been
specified, one can solve~\eqref{psi_evol} for ${}^{(2)}\Psi$ and then
successively use~\eqref{gov_scalar1},~\eqref{matter} and~\eqref{gov_scalar4}
to calculate ${}^{(2)}\Phi, {}^{(2)}\Delta_{\mathrm p}$ and
${}^{(2)}V_{\mathrm p}$, respectively.

\subsubsection*{The uniform curvature form}

To obtain the uniform curvature form we substitute~\eqref{leadein_curv}
and~\eqref{Tmode} into~\eqref{ein3} and then apply the mode extraction
operators. For the sake of brevity we give only the scalar mode since the
vector and tensor modes have essentially the Poisson form. \vspace{5mm}

\emph{Scalar mode}
\begin{subequations} \label{scalar_eq_curv}
\begin{align}
{\mathcal L}_B{}^{(2)}{\bf B} + {}^{(2)}{\bf A} &=\,\, - {}^{(2)}\Pi_{\mathrm c} +
{\mathcal S}^{ij}\, \hat{\mathcal{G}}{}_{ij}({\bf f}_{\mathrm c},{\bf f}_{\mathrm c}),   \,\label{bfB_evol}\\
\cH\!\left({\mathcal L}_A{}^{(2)}{\bf A} + {\cal C}_G^2 {\bf D}^2\,{}^{(2)}{\bf B}\right)
&=\,\, \sfrac12{}^{(2)}\Gamma_{\mathrm c} +
 \sfrac13{\bf D}^2[{}^{(2)}\Pi_{\mathrm c}- {\mathcal S}^{ij}\, \hat{\mathcal{G}}{}_{ij}({\bf f}_{\mathrm c},{\bf f}_{\mathrm c}) ]  -
\sfrac12\mathcal{G}({\bf f}_{\mathrm c},{\bf f}_{\mathrm c}), \label{Phicurv_evol}\\
{\cal H}\!\left({\bf D}^2 + 3K\right){}^{(2)}{\bf B} &=\,\, -
\sfrac12{}^{(2)}\Delta_{\mathrm c} +
 \sfrac12{\cal
S}^i\,\mathcal{G}_i({\bf f}_{\mathrm c},{\bf f}_{\mathrm c}) , \label{Poisson_C} \\
\cH {}^{(2)}{\bf A} - K{}^{(2)}{\bf B} &=\,\, - \sfrac12 {}^{(2)}V_{\mathrm c} +
\sfrac12{\cal
S}^i\,\mathcal{G}^0\!_i({\bf f}_{\mathrm c},{\bf f}_{\mathrm c}). \label{V_F}
\end{align}
\end{subequations}
Here ${}^{(2)}\bf A$ and ${}^{(2)}\bf B$ are the second order uniform curvature metric
gauge invariants and ${\bf f}_{\mathrm c}$ is shorthand for the first order (in time)
metric metric perturbation ${\bf f}_{ab}[X_{\mathrm c}]$ as given by~\eqref{bf_curv}.

These equations differ from the Poisson form~\eqref{gov_scalar} in that the
evolution of the scalar potentials ${\bf A}$ and ${\bf B}$ is governed by two
coupled first order partial differential equations~\eqref{bfB_evol}
and~\eqref{Phicurv_evol}.  In order to obtain a solution one has to follow a two step
procedure, as with the Poisson form.

\subsubsection*{Commentary}

The systems of equations~\eqref{gov_scalar}
and~\eqref{scalar_eq_curv}, together with the
expressions~\eqref{Einstein_source1}--\eqref{Q_sf} for the source terms,
are new and constitute one of the main results of this paper.
Either set of equations determines the behaviour of second order scalar perturbations
of an FL cosmology with arbitrary stress-energy content.
 We emphasize that the specific form of the
evolution equations~\eqref{gov_scalar}
and~\eqref{scalar_eq_curv} depends on the matter terms, specifically,
the $\Gamma$-terms and the $\Pi$-terms, both at first order and at
second order. These quantities are determined by
the stress-energy tensor using equation~\eqref{T_gi} in conjunction with~\eqref{Gamma_mode}
and~\eqref{Tdecomp1}. In the next section we illustrate how to
calculate these quantities for the simple case of a
perfect fluid.

Equations~\eqref{gov_scalar} and~\eqref{scalar_eq_curv}
illustrate a fundamental difference between second order and first order perturbations.
The analysis of linear perturbations is simplified by the fact that the three
modes, namely scalar, vector and tensor, decouple and hence can be analyzed
separately. At second order each mode involves leading order terms, described
by the same operators that occur at first order, but also complicated
quadratic source terms that, in general, contain all linear modes, which
leads to a phenomenon one may refer to as \emph{source mode coupling}. For example,  this means
that the first order metric perturbation ${\bf f}_{\mathrm p}$ that determines the
Einstein source terms in~\eqref{gov_scalar} contains scalar, vector and tensor
modes  in general  (see~\eqref{bf_Poisson}). In other words the vector and tensor perturbations at linear order
contribute to the scalar perturbation at second order.
On the other hand equations~\eqref{gov_scalar} show that a purely scalar linear perturbation ({\it i.e.} if the vector and
tensor modes at the linear level are assumed to be zero, as is often done) generate all three
modes at second order.\footnote{See, for example, Bartolo {\it et al}
(2004a), section 3.1.}

\subsection{Matter gauge invariants for a perfect fluid and $\Lambda$}

In this section we determine the second order stress-energy perturbations
${}^{(2)}\!\Gamma[X]$, ${}^{(2)}\!\Pi[X]$, ${}^{(2)}\!\Pi_i[X]$ and
${}^{(2)}\!\Pi_{ij}[X]$ that appear in the governing equations in Poisson
form, as given by equations~\eqref{gov_scalar}, \eqref{gov_vector}
and~\eqref{gov_tensor}, when the stress-energy tensor describes a perfect
fluid and a cosmological constant $\Lambda$:
\begin{equation}\label{pf}
T^a\!_b = \left(\rho + p\right)\!u^a u_b + p\delta^a\!_b - \Lambda\delta^a\!_b.
\end{equation}
Here $u^a$, $\rho$, $p$ are the fluid's 4-velocity, energy-density and
pressure, respectively. Since $\Lambda$ is $\epsilon$-independent it follows
that it does not appear in ${}^{(1)}\!T^a\!_b$ and ${}^{(2)}\!T^a\!_b$, and
it only affects the perturbed field equations indirectly via background
quantities determined by the zeroth order field equations. For simplicity we
assume $p=w\rho$ with $w=constant$, which implies that
\begin{equation}
{}^{(r)}p={\mathcal C}_T^2{}^{(r)}\!\rho, \quad
\text{with} \quad {\mathcal C}_T^2=w, \quad \text{for}\quad r=1,2.
\end{equation}

In order to find the desired quantities we need to calculate
${}^{(2)}\!\Gamma[X]$ and ${}^{(2)}\!{\hat{\bf T}}_{ij}[X]$ for the
stress-energy tensor \eqref{pf}. We begin by expanding~\eqref{pf} to second
order, which yields
\begin{subequations}\label{diag_T}
\begin{align}
a^2\,{}^{(2)}\!T^0\!_0 &= -a^2\,{}^{(2)}\!\rho - 2{\cal
A}_T\gamma^{ij}{}^{(1)}\!v_i \left({}^{(1)}\!v_j -
f_{0j}\right),\\
a^2\,{}^{(2)}\!T^k\!_k &= 3a^2\,{}^{(2)}\!p + 2{\cal
A}_T\gamma^{ij}{}^{(1)}\!v_i \left({}^{(1)}\!v_j -
f_{0j}\right),\\
a^2{}^{(2)}\hat{T}_{ij} &= 2{\cal
A}_T\!\left({}^{(1)}\!v_{\la i} - f_{0\la i}\right)\!{}^{(1)}\!v_{j\ra},
\end{align}
\end{subequations}
where ${}^{(1)}v_i :=a^{-1}{}^{(1)}u_i$. Applying the Replacement
Principle\footnote{To obtain ${}^{(2)}\!{\bf T}^a\!_b[X]$ replace each
perturbation variable on the right side by its gauge invariant formed by
$X$-compensation {\it i.e.} ${}^{(2)}\!\rho\rightarrow{}^{(2)}\!\brho[X]$ and
${\mathcal A}_T{}^{(1)}\!v_i\rightarrow{}^{(1)}\!V_i[X]$. See
equation~\eqref{T_RP}.} gives
\begin{subequations}
\begin{align}
{}^{(2)}\!{\bf T}^0\!_0[X] &= -{}^{(2)}\!\brho[X] - {\cal U}^k\!_k[X],   \label{^(2)T_bf} \\
{}^{(2)}\!{\bf T}^k\!_k[X] &= - 3{\mathcal
C}_T^2{}^{(2)}\!\brho[X] + {\cal U}^k\!_k[X], \label{^(2)T_bf2} \\
{}^{(2)}\!{\hat{\bf T}}_{ij}[X] &= \hat{\cal U}_{ij}[X],
\end{align}
\end{subequations}
where
\begin{equation}
{\cal U}_{ij}[X] := 2{\cal A}_T^{-1} {}^{(1)}\!V_i[X]
\left( {}^{(1)}\!V_j[X] - {\mathcal A}_T{\bf f}_{0j}[X]\right),
\end{equation}
and
\begin{equation}
{\cal A}_T := a^2({}^{(0)}\!\rho + {}^{(0)}\!p), \qquad
{}^{(1)}\!V_i[X]:= {}^{(1)}\!{\bf T}^0\!_i[X].
\end{equation}
Equations \eqref{^(2)T_bf} and \eqref{^(2)T_bf2},
together with~\eqref{gamma}, result in
\begin{equation}
{}^{(2)}\!\Gamma[X] =  \sfrac23(1-3{\mathcal C}_T^2){\cal U}^k\!_k[X].
\end{equation}
As expected, we see that $^{(2)}\!\Gamma[X]$ and ${}^{(2)}\!{\hat{\bf
T}}_{ij}[X]$ are purely source terms.\footnote{Since $^{(1)}\!\Gamma[X]=0$
and ${}^{(1)}\!{\hat{\bf T}}_{ij}[X]=0$ for a perfect fluid it follows that
the leading order term in $^{(2)}\!\Gamma[X]$ and in ${}^{(2)}\!{\hat{\bf
T}}_{ij}[X]$ is zero.}

For simplicity we now assume that the linear vector modes $\tilde{V}_i[X]$ and
${\bf B}_i[X]$ are zero, i.e.,
\begin{equation}
V_i[X] = {\bf D}_i V[X], \qquad {\bf f}_{0j}[X] = {\bf D}_i
{\bf B}[X].
\end{equation}
We then choose $X=X_{\mathrm p}$, noting that ${\bf
B}[X_{\mathrm p}]=0$ and $V[X_{\mathrm p}]\equiv V$.
It follows that
\begin{equation}
{\cal U}_{ij}[X_{\mathrm p}] = 2{\cal A}_T^{-1}({\bf D}_i V)({\bf D}_j V),
\end{equation}
and hence
\begin{equation}  \label{simple_gamma}
{}^{(2)}\!\Gamma[X]=
\sfrac23(1-3{\mathcal C}_T^2){\cal A}_T^{-1}({\bf D}V)^2, \qquad
{}^{(2)}\!{\hat{\bf T}}_{ij}[X_{\mathrm p}] = 2{\cal
A}_T^{-1}{\bf D}_{\la i}V\,{\bf D}_{j\ra}V.
\end{equation}
The required quantities ${}^{(2)}\!\Pi[X_{\mathrm p}]$,
${}^{(2)}\!\Pi_i[X_{\mathrm p}]$ and ${}^{(2)}\!\Pi_{ij}[X_{\mathrm p}]$ are
then obtained by applying the mode extraction operators as
in~\eqref{Tdecomp1}.

In order to facilitate comparison with the literature we relate the gauge
invariants associated with the matter density and with the matter velocity in
the Poisson gauge at second order, denoted ${}^{(2)}\!{\brho}_{\mathrm p}$,
and ${}^{(2)}\!{\bf v}_{\mathrm p}$, respectively, to our stress-energy gauge
invariants ${}^{(2)}\!{\Delta}_{\mathrm p}$ and ${}^{(2)}\!{V}_{\mathrm p}$:
\begin{subequations}
\begin{align}
{}^{(2)}\!{\brho}_{\mathrm p}= &{}^{(2)}\!{\Delta}_{\mathrm p}+
3{\mathcal H}{}^{(2)}\!{V}_{\mathrm p}- 2{\mathcal A}_T^{-1}({\bf D}V_{\mathrm p})^2, \label{rho_p}  \\
{}^{(2)}\!{\bf v}_{\mathrm p} =& {\mathcal A}_T^{-1}\left( {}^{(2)}\!{V}_{\mathrm p} +
2{\mathcal S}^i\left[\left(\Psi - \frac{{\brho}_{\mathrm p}}{3\Omega_m {\mathcal H}^2}\right)
{\bf D}_i V_{\mathrm p}\right] \right), \label{v_p}
\end{align}
where
\be {\brho}_{\mathrm p}= \Delta+3{\mathcal H}{V}_{\mathrm p},
\qquad \Omega_m=a^2\,{}^{(0)}\!\rho/(3{\cal H}^2). \ee
\end{subequations}
%

\subsection{The governing equations in a simple example}

We now specialize the general governing equations~\eqref{gov_scalar},
\eqref{gov_vector} and~\eqref{gov_tensor} at second order in the Poisson
gauge to a perfect fluid and a cosmological constant with only scalar first
order contributions and $\Phi=\Psi$, {\it i.e.} we impose the metric
assumptions \eqref{m_assump}. We hence substitute for the metric source terms
${\cal G}({\bf f}_{\mathrm p},{\bf f}_{\mathrm p})$
from~\eqref{special_source} and for the stress-energy perturbations
${}^{(2)}\Gamma$ and the three ${}^{(2)}\Pi$-terms from~\eqref{Tdecomp1}
and~\eqref{simple_gamma}. This leads to the following governing equations:
\vspace{3mm}

\emph{Scalar mode}
\begin{subequations} \label{gov_scalar3}
\begin{align}
{}^{(2)}\!\Psi - {}^{(2)}\!\Phi &= - 4\Psi^2 +2{\cal S}^{ij} {\cal M}_{ij}(\Psi,\Psi),  \label{psi-phi}\\
\begin{split}
\left({\cal L} - {\cal C}_G^2{\bf D}^2\right)\!{}^{(2)}\!\Psi
&= \sfrac13(1-3{\mathcal C}_T^2){\cal A}_T^{-1}({\bf D}V)^2
+ \sfrac16\!\left(1 + 3{\cal C}_G^2\right){\mathcal R}(\Psi,\Psi)\\
& + \sfrac43({\bf D}\Psi)^2 - \sfrac43{\bf D}^2\Psi^2
+ 2\left({\cal H}{\cal L}_A + \sfrac13{\bf D}^2\right){\cal S}^{ij} {\cal M}_{ij}(\Psi,\Psi),
\end{split} \label{psi_evol3} \\
({\bf D}^2 + 3K){}^{(2)}\!\Psi &= \sfrac12 {}^{(2)}\!\Delta
- \sfrac12{\mathcal R}(\Psi,\Psi)
- 6{\cal H}{\cal S}^i\!\left((\partial_\eta\Psi)({\bf D}_i\Psi)\right) ,
\label{matter2}\\
(\partial_{\eta} + {\cal H}){}^{(2)}\!\Psi &=
- \sfrac12{}^{(2)}\!V -
2{\cal S}^i((\partial_\eta\Psi)({\bf D}_i\Psi))  + 2{\cal H}{\cal S}^{ij} {\cal M}_{ij}(\Psi,\Psi),  \label{V2}
\end{align}
where
\begin{align}
{\mathcal R}(\Psi,\Psi) &:= 2\left[ 3(\partial_\eta \Psi)^2  -
 5({\bf D}\Psi)^2 +
4\!\left( {\bf D}^2 + 3K\right)\!\Psi^2\right],\\
{\cal M}_{ij}(\Psi,\Psi) &:= 2({\bf D}_{\la i}\Psi)({\bf D}_{j\ra}\Psi) +
{\cal A}_T^{-1}({\bf D}_{\la i}V)({\bf D}_{j\ra}V),  \label{V_ij}
\end{align}
and $V$ is given by\footnote{This is obtained from the governing equations
for the first order perturbations.}
\begin{equation}  \label{V_evol}
V = -2(\partial_\eta + {\cal H})\Psi.
\end{equation}
\end{subequations}

\emph{Vector mode}
\begin{subequations}\label{vector_eq3}
\begin{align}
{\cal L}_B{}^{(2)}\!{\bf B}_i &= -4{\cal V}_i\!^{jk}{\cal M}_{jk}(\Psi,\Psi),  \label{vector_evolsc}\\
({\bf D}^2 + 2K){}^{(2)}\!{\bf B}_i &=  2{}^{(2)}\!{V}_i +
8{\cal V}_i\!^j((\partial_\eta\Psi)({\bf D}_j\Psi)). \label{vector_V2}
\end{align}
\end{subequations}

\emph{Tensor mode}
\begin{equation}\label{tensor_evol3}
\left({\cal L}_B\partial_\eta + 2K - {\bf D}^2\right){}^{(2)}\!{\bf C}_{ij} =
2{\cal T}_{ij}\!^{km}{\cal M}_{km}(\Psi,\Psi).
\end{equation}

\subsubsection*{Commentary}

An attractive feature of the Poisson gauge at first order is that if the
anisotropic stress is zero (for example, for a perfect fluid or a scalar
field) then $\Phi=\Psi$, {\it i.e.} there is only one metric gauge invariant
for the scalar mode. Equation~\eqref{psi-phi} shows that this feature is not
preserved at second order due to the presence of the source terms. There are
thus two second order metric gauge invariants,
${}^{(2)}\Phi\neq{}^{(2)}\Psi$. We shall refer to ${}^{(2)}\Phi$ as the {\it
second order Bardeen potential} and  to ${}^{(2)}\Psi$ as the {\it second
order Bardeen curvature}, to distinguish their roles.\footnote{Strictly
speaking one should also make this distinction at first order, but since
$\Phi=\Psi$ in many applications we simply refer to $\Psi$ as the Bardeen
potential.}

The system of equations~\eqref{gov_scalar3}-\eqref{tensor_evol3} is closely
related to but simpler than equations that appear in the literature.
 Nakamura (2007) has given a system of
equations that can be transformed into\footnote
{His equations (6.38), (6.44),
(6.41) and (6.42) can be used to obtain the first four of our
equations~\eqref{gov_scalar3}, his equations (6.39) and (6.33) can be transformed into
our equations~\eqref{vector_eq3}, and finally his equation (6.40) yields  our
equation~\eqref{tensor_evol3}. We also refer to Nakamura (2006) for a
brief summary.}
our equations~\eqref{gov_scalar3}-\eqref{tensor_evol3}. There are, however two
important differences. First, Nakamura chooses the Bardeen potential
${}^{(2)}\!\Phi$ rather than the Bardeen curvature ${}^{(2)}\!\Psi$ to be the
primary metric gauge invariant for the scalar mode,  which has a major
drawback: the source terms contain second order time derivatives
and are significantly more complicated. The second
difference is in the treatment of the density perturbation. We use
${}^{(2)}\!\Delta_{\mathrm p}$, which satisfies a generalized Poisson
equation and can be viewed as being analogous at second order to the well
known Bardeen gauge invariant $\epsilon_m$. In contrast, Nakamura uses the
gauge invariant ${}^{(2)}\!\brho_{\mathrm p}$ associated with the density
perturbation at second order in the Poisson gauge, as given
by~\eqref{rho_p}.\footnote {Note that ${}^{(2)}\!\brho[X_{\mathrm p}]\equiv
a^2\,{}^{(2)}\!\varepsilon$ in Nakamura's notation.}

Equations~\eqref{vector_eq3} and~\eqref{tensor_evol3} show that a purely
scalar linear perturbation gives rise to a vector and a tensor perturbation
at second order, with the link provided by the tensor ${\mathcal M}_{ij}$.
The gravitational waves described by this tensor perturbation have been
investigated in detail by  Ananda {\it et al} (2007) and Baumann {\it et al}
(2007).

\section{Discussion}\label{sec:disc}

The systems of equations~\eqref{gov_scalar}
and~\eqref{scalar_eq_curv} which govern nonlinear perturbations, together with the
expressions~\eqref{Einstein_source1}--\eqref{Q_sf} for the source terms,
are new and constitute one of the main results of this paper.
Because of their generality these equations
provide a starting point for determining the behaviour of nonlinear
perturbations of FL cosmologies with any given stress-energy content,
using either the Poisson gauge or the uniform curvature gauge.\footnote
{ For example, by specializing these equations
we can derive in an efficient manner the various equations for second order perturbations
that appear in the papers by Bartolo and collaborators (Poisson form)
and by Malik and collaborators (uniform curvature form),
referred to in the introduction.}
These equations exhibit the same concise structure as the governing equations for
linear perturbations given in UW1 (see equations (52) and (54)), in which the evolution of the
metric perturbations is determined by the second order factored differential operator
${\cal L}$ in the Poisson form and by the pair of first order differential operators
${\cal L}_A$ and ${\cal L}_B$ in the uniform curvature form. This structure arises directly
from our use of specific linear combinations of the
components of the perturbed Einstein tensor and their
derivatives, as given by~\eqref{ein_minimal} (see UW1, equations
(39) and (40) for the motivation). Indeed the three operators are visible
at an early stage in the derivation in the leading order expressions~\eqref{ein_leading}
for the perturbed Einstein tensor. This is in contrast to the literature where it is customary to simply calculate all the components
$^{(2)}G^a\!_b$ of the perturbed Einstein tensor, and then form linear
combinations of the perturbed Einstein equations (see for example,
Acquaviva {\it et al} (2003), equation (4) for the perturbed metric and
Appendix A5 for the perturbed Einstein tensor, and Bartolo {\it
et al} (2004a), equations (104) and (A.36)--(A.43).).
This process involves more extensive
calculations than in our approach and may lead to expressions that are not optimally simplified
while hiding important mathematical structures.

There is one issue that deserves particular attention, namely the fact that
there is no unique choice of gauge invariant associated with the
perturbations of the matter density. At the linear level there are three
commonly used choices, the Poisson gauge invariant, the uniform curvature
gauge invariant that is related to one of the so-called conserved
quantities\footnote{See, for example, Malik and Wands (2004), equations
(4.17) and (4.27).}, and the total matter (or comoving) gauge invariant. The
last-mentioned is the well-known Bardeen gauge invariant, and is related to
the spatial gradient of the matter density orthogonal to the fluid flow. At
the second order level the situation is more complicated and requires further
investigation.\footnote{See Bartolo {\it et al} (2010a), section 3 and in
particular equation (29) for ${}^{(2)}\!{\brho}_{\mathrm p}$ in a $\Lambda
CDM$ model, Christopherson and Malik (2009), equation (4.8) for
${}^{(2)}\!{\brho}_{\mathrm c}$ and Noh and Huang (2004), equation (273) for
${}^{(2)}\!{\brho}_{\mathrm v}$.} The appropriate choice may depend on the
physical situation under consideration.

In this paper we have focussed exclusively on using the perturbed Einstein
field equations to describe the dynamics of nonlinear perturbations. There
are, however, two alternatives to the direct use of the Einstein equations.
First, one can use the perturbed conservation equations for the stress-energy
tensor,\footnote {See, for example Bartolo {\it et al} (2004b), equation
(4.3), and Noh and Hwang (2004), equations (104) and (200).}
 and second, one can use the $1+3$
 formalism,\footnote{See Bruni {\it et al} (1992) for a
comprehensive treatment of linear perturbations using this formalism.}
expanding the exact equations to second order and making them
gauge-invariant. More work needs to be done in this regard. An additional
aspect of the dynamics of scalar perturbations that we have likewise not
touched on is that under certain conditions ({\it i.e.} in the long
wavelength regime) the governing equations admit so-called conserved
quantities, {\it i.e.} quantities that remain approximately constant during a
restricted epoch. These quantities, which were initially introduced for
linear perturbations (see, for example UW2 section 4 for a unified overview)
have now been generalized to second order perturbations.\footnote {Malik and
Wands (2004), equations (4.17) and (4.18) and Christopherson and Malik
(2009), equations (4.11)--(4.13).}

\subsection*{Acknowledgments}
CU is supported by the Swedish Research Council (VR grant 621-2009-4163). CU
also thanks the Department of Applied Mathematics at the University of
Waterloo for kind hospitality. JW acknowledges financial support from the
University of Waterloo.


\begin{appendix}

\section{Derivation of the perturbation equations}\label{app:der}

\subsection{Exact curvature expressions}\label{app:exactcurv}

In UW1 we derived an exact expression for the Riemann curvature tensor
$R^{ab}\!_{cd}(\epsilon)$ of $g_{ab}(\epsilon)=a^2{\bar g}_{ab}(\epsilon)$ by
replacing the covariant derivative ${}^\epsilon\bna\!_a$ of
$g_{ab}(\epsilon)$ with the covariant derivative ${}^0\!\bar{\bna}\!_a$ of
$\bar{g}_{ab}(0)=\gamma_{ab}$. We first make the following definitions:
\begin{subequations}
\begin{gather}
r_a:={}^0\!\bar{\bna}\!_a (\ln a), \label{r_a} \\
\tilde{Q}^a\!_{bc}(\epsilon)
:= \bar{g}^{ad}(\epsilon)
\left({}^0\!\bar{\bna}\!_{(b}\,\bar{g}_{c)d}(\epsilon)
- \sfrac{1}{2}\,{}^0\!\bar{\bna}\!_{d}\,\bar{g}_{bc}(\epsilon)\right). \label{Qcosmo3}
\end{gather}
\end{subequations}
Since ${}^0\!\bar{\bna}\!_a \gamma_{bc}=0$ it follows that
\be \label{Q_0} \tilde{Q}^a\!_{bc}(0) =0. \ee
The desired expression is
as follows (see UW1, (B.8), (B.10) and (B.12b)):
\begin{subequations}\label{Riemeps}
\be
a^2R^{ab}\!_{cd}(\epsilon) =
\bar{R}^{ab}\!_{cd}(\epsilon) + 4\delta^{[a}\!_{[c}
\bar{U}^{b]}\!_{d]}(\epsilon),\label{Riemeps2}  \ee
where
\begin{align}
\bar{R}^{ab}\!_{cd}(\epsilon) &=
\bar{g}^{be}(\epsilon)\left( {}^0\!\bar{R}^{a}\!_{ecd} + 2{}^0\!\bar{\bna}\!_{[c }\tilde{Q}^a\!_{d]e}(\epsilon) +
2\tilde{Q}^a\!_{f[c}(\epsilon)\tilde{Q}^f\!_{d]e}(\epsilon)\right),\label{Riemdowneps}\\
\bar{U}^{b}\!_{d}(\epsilon) &= -
\left[\bar{g}^{be}(\epsilon)\,({}^0\!\bar{\bna}\!_{d} - r_{d}) +
\sfrac{1}{2}\delta^{b}\!_{d}\,\bar{g}^{ef}(\epsilon)\,r_f -
\bar{g}^{bf}(\epsilon)\,\tilde{Q}^e\!_{df}(\epsilon) \right]\!r_e. \label{Ueps}
\end{align}
\end{subequations}
Here $\bar{R}^{ab}\!_{cd}(\epsilon)$ is the curvature tensor of the metric
$\bar{g}_{ab}(\epsilon)$, and ${}^0\!\bar{R}^a{}_{bcd}$ is the curvature
tensor of the metric $\gamma_{ab}$. Note that $^0\!\bar{R}^{ab}\!_{cd} =
\gamma^{be}\,{}^0\!\bar{R}^{a}\!_{ecd}$ is zero if at least one index is
temporal, while if all indices are spatial
\be\label{threecurv} {}^0\!\bar{R}^{ij}\!_{km} =
2K\delta^{[i}\!_{[k} \delta^{j]}\!_{m]}. \ee
%

\subsection{First and second order gauge-variant perturbations}\label{app:curvperturb}

Our first goal is to derive expressions for the geometric operators
$\mathsf{R}^{ab}\!_{cd}({}^{(r)}\!f)$ and ${\cal R}^{ab}\!_{cd}(f,f)$ that
determine the perturbed Riemann tensor through
equations~\eqref{rie_operator}. To accomplish this we need the Taylor
expansion of $R^{ab}\!_{cd}(\epsilon)$, given by~\eqref{Riemeps}.

We begin by deriving expressions for  ${}^{(1)}\!\bar{g}^{ab}$ and
${}^{(2)}\!\bar{g}^{ab}$. We Taylor expand the relation $\delta^a\!_b =
\bar{g}^{ac}(\epsilon)\,\bar{g}_{cb}(\epsilon)$, which gives
\be  \label{gcontn}
{}^{(1)}\!\bar{g}^{ab} = -f^{ab},\qquad
{}^{(2)}\!\bar{g}^{ab} = - {}^{(2)}\!f^{ab} + 2f^{ac}f_c\!^{b}, \ee
where we used the following relations
\begin{subequations}\label{Leibniz}
\begin{align}
{}^{(1)}\!(AB) &= {}^{(0)}\!A\,{}^{(1)}\!B + {}^{(1)}\!A\,{}^{(0)}\!B,\\
{}^{(2)}\!(AB) &= {}^{(0)}\!A\,{}^{(2)}\!B + 2\,{}^{(1)}\!A\,{}^{(1)}\!B +
{}^{(2)}\!A\,{}^{(0)}\!B.
\end{align}
\end{subequations}
We next Taylor expand $\tilde{Q}^a\!_{bc}(\epsilon)$, as
given by~\eqref{Qcosmo3}, and use~\eqref{Leibniz}. This leads to
\begin{subequations}
\be \label{Q_tilde}
{}^{(1)}\!\tilde{Q}^a\!_{bc} =
\tilde{\mathsf{Q}}^a\!_{bc}(f),\qquad
{}^{(2)}\!\tilde{Q}^a\!_{bc} =
\tilde{\mathsf{Q}}^a\!_{bc}\left({}^{(2)}\!f\right) - 2f^{ad}\tilde{\mathsf{Q}}_{dbc}(f).  \ee
where the operator $\tilde{\mathsf{Q}}^a\!_{bc}$ is defined by
\be \label{Q_op}  \tilde{\mathsf{Q}}^a\!_{bc}(f): = \gamma^{ad}({}^0\!\bar{\bna}\!_{(b}f_{c)d} -
\sfrac12\,{}^0\!\bar{\bna}\!_d f_{bc}) . \ee
\end{subequations}
Finally Taylor expanding $R^{ab}\!_{cd}(\epsilon)$ as given
by~\eqref{Riemeps} and using~\eqref{rie_operator}, in conjunction
with~\eqref{gcontn},~\eqref{Q_0} and~\eqref{Q_tilde}, gives the leading order
term
\begin{subequations}\label{Rie_leading}
\be
\mathsf{R}^{ab}\!_{cd}({}^{(r)}\!f) = \bar{\mathsf{R}}^{ab}\!_{cd}({}^{(r)}\!f)
+ 4\delta^{[a}\!_{[c} \bar{\mathsf{U}}^{b]}\!_{d]}({}^{(r)}\!f),\quad r=1,2,    \label{mastercurv2a} \ee
where
\begin{align}
\bar{\mathsf{R}}^{ab}\!_{cd}({}^{(r)}\!f) &=
-2\,{}^0\!\bar{\bna}\!_{[c}{}^0\!\bar{\bna}^{[a}\,{}^{(r)}\!f_{d]}\!^{b]}
+ {}^{(r)}\!f_e\!^{[a}\,{}^0\!\bar{R}^{b]e}\!_{cd},\label{barcurvfirst}\\
\bar{\mathsf{U}}^{b}\!_{d}({}^{(r)}\!f) &= {\mathsf{U}}^{be}\!_{de}({}^{(r)}\!f), \label{barUfirst}
\end{align}
\end{subequations}
and the source term
\begin{subequations}  \label{Rie_source}
\be
{{\cal R}}^{ab}\!_{cd}(f,f) =
\bar{{\cal R}}^{ab}\!_{cd}(f,f) +
4\delta^{[a}\!_{[c} \bar{{\cal U}}^{b]}\!_{d]}(f,f),  \label{mastercurv2b} \ee
where
\begin{align}
\bar{\mathcal{R}}^{ab}\!_{cd}(f,f) &=
2f_e\!^{[a}\!\left(2\bar{\mathsf{R}}^{b]e}\!_{cd}(f) -
f_f\!^{b]}\,{}^0\!\bar{R}^{ef}\!_{cd}\right)
- 4\tilde{\mathsf{Q}}^{f[a}\!_{[c}(f)\tilde{\mathsf{Q}}_{|f|}\!^{b]}\!_{d]}(f), \label{barcurvsecond}\\
\bar{{\cal U}}^{b}\!_{d}(f,f) &=
-2f_e\!^f\!\left({\mathsf{U}}^{be}\!_{df}(f)  +
\delta^{b}\!_f\,\tilde{\mathsf{Q}}^{ge}\!_{d}(f)\,r_g\right)\!.  \label{barUsecond}
\end{align}
\end{subequations}
In the above equations\footnote{To obtain~\eqref{barcurvfirst}
we used
${}^0\!\bar{\bna}_{[c}{}^0\!\bar{\bna}_{d]}{}^{(r)}\!f^{ab} =
{}^{(r)}\!f_e\!^{(a}\,{}^0\!\bar{R}^{b)e}\!_{cd}$, while
${}^0\!\bar{\bna}_c {}^{(r)}\!f^{ab} =
2\tilde{\mathsf{Q}}^{(ab)}\!_c({}^{(r)}\!f)$ was used to
obtain~\eqref{barcurvsecond}.} we have also defined
\begin{equation}
{\mathsf{U}}^{be}\!_{df}({}^{(r)}\!f) :=
\left[{}^{(r)}\!f^{be}\left({}^0\!\bar{\bna}\!_{d} - r_{d}\right) +
\sfrac{1}{2}\delta^{b}\!_{d}\,{}^{(r)}\!f^{eh}\,r_h +
\gamma^{bg}\,\,\tilde{\mathsf{Q}}^e\!_{dg}({}^{(r)}\!f)\right]\!r_f .
\end{equation}
Equations~\eqref{Rie_leading} and~\eqref{Rie_source} constitute one of the main results of this paper.
They express the first and second order Riemann tensor perturbations,
as given by~\eqref{rie_operator}, in terms of the metric perturbations
as given by equations~\eqref{metricperturb}, with the second order perturbation written
as the sum of a leading order term and a source term.

To proceed further we need to introduce local coordinates as in
section~\ref{sec:coords}, which implies that the covariant derivative of a
tensor, ${}^0\!\bar{\bna}\!_a A$, and the gradient $r_a$ defined
by~\eqref{r_a}, assume the form
\begin{equation} \label{deriv} {}^0\!\bar{\bna}\!_0 \, A = \partial_\eta A, \qquad
{}^0\!\bar{\bna}\!_i A = {\bf D}_i A , \qquad r_0 ={\cal H}, \qquad r_i=0.
\end{equation}
where $\partial_\eta$ denotes partial differentiation with respect to $\eta$
and ${\bf D}_i$ is the spatial covariant derivative of $\gamma_{ij}$.

We have already used equations~\eqref{Rie_leading}  to write the Einstein
leading terms in the form~\eqref{ein_leading} in the main text. We now use
equations~\eqref{Rie_source} to write the Einstein source terms, as given
by~\eqref{source_G}, in the form
\begin{subequations}  \label{Einstein_source1}
\begin{align}
\hat{\mathcal{G}}_{ij}(f,f) &=
\hat{{\mathcal{S}}}_{ij} + \hat{{\mathcal{W}}}_{ij} +
2\tilde{\mathsf{Q}}_{\alpha\beta\la i} \tilde{\mathsf{Q}}^{\alpha\beta}\!_{j\ra}
- 2\tilde{\mathsf{Q}}^{\alpha\beta}\!_\beta \tilde{\mathsf{Q}}_{\alpha\la ij\ra},\\
{\mathcal{G}}(f,f) &= - \sfrac13\!\left[(1 + 3{\cal C}_G^2){\mathcal R}
+ 2{\mathcal{R}}_0 + {\mathcal{W}}^k\!_k
- 6{\cal H}{\cal L}_A(f_{0\alpha}f_0\!^\alpha)\right],\label{Ggen}\\
{\mathcal{G}}_{i}(f,f) &= {\bf D}_i{\mathcal R} - 3{\cal H}\bar{\mathcal{G}}^{0}\!_{i},\\
{\mathcal{G}}^0\!_{i}(f,f) &= - 2{\cal H}{\bf D}_i(f_{0\alpha}f_0\!^\alpha)+ \bar{\mathcal{G}}^{0}\!_{i},
\end{align}
\end{subequations}
where
\begin{subequations} \label{Einstein_source2}
\begin{align}
{\cal W}_{ij}(f,f) &:= 4{\cal H}\!\left[
f_{0\alpha}\tilde{\mathsf{Q}}^{\alpha}\!_{ij} - \sfrac12 f_{0i}{\bf D}_{j}f_{00} -
f_{ik}\tilde{\mathsf{Q}}^{0k}\!_{j}\right],\\
\hat{\mathcal{S}}_{ij}(f,f) &:= -2\!\left[\bar{\mathsf{R}}^{\alpha k}\!_{\beta \la i}\gamma_{j\ra k} f_\alpha\!^\beta
+ \bar{\mathsf{R}}^{\alpha\beta}\!_{\alpha\la i} f_{j\ra \beta} - K(\hat{f}^k\!_{\la i}\hat{f}_{j\ra k} - \sfrac13 f^k\!_k \hat{f}_{ij})\right],\\
{\mathcal R}(f,f) &:=  {\mathcal{W}}^k\!_k       -2f_\alpha\!^k\bar{\mathsf{R}}^{\alpha m}\!_{km} +
K(\hat{f}^k\!_m\hat{f}^m\!_k - \sfrac23 (f ^k\!_k)^2) - 2\tilde{\mathsf{Q}}^{\alpha[k}\!_k \tilde{\mathsf{Q}}_\alpha\!^{m]}\!_m,\\
\bar{\mathcal{G}}^{0}\!_{i}(f,f) &\: = {\mathcal{R}}_i + 2K(\sfrac23  f_{0i}f^k\!_k - f_{0m}\hat{f}^m\!_i),\\
{\mathcal{R}}_\alpha(f,f) &:= 2\!\left[(f_{00} - \sfrac13 f ^k\!_k)\bar{\mathsf{R}}^{0m}\!_{\alpha m}
+ f_{0k}\bar{\mathsf{R}}^{km}\!_{\alpha m} -
\hat{f}^m\!_k\bar{\mathsf{R}}^{0k}\!_{\alpha m}\right] - 4\tilde{\mathsf{Q}}^{\beta[0}\!_\alpha \tilde{\mathsf{Q}}_\beta\!^{m]}\!_m.
\end{align}
\end{subequations}
For notational brevity we have dropped the arguments $(f,f)$ of the
quantities~\eqref{Einstein_source2} and the argument $f$ of
$\tilde{\mathsf{Q}}^{a}\!_{bc}(f)$ and $\bar{\mathsf{R}}^{ab}\!_{cd}(f)$ when
they appear on the right side of~\eqref{Einstein_source1}
and�~\eqref{Einstein_source2}. It remains to give the components of
$\bar{\mathsf{R}}^{ab}\!_{cd}(f)$ and $\tilde{\mathsf{Q}}^a\!_{bc}(f)$, that
are defined by~\eqref{barcurvfirst} and~\eqref{Q_op}:
\begin{subequations}\label{R_sf}
\begin{align}
\bar{\mathsf{R}}^{0j}\!_{0m} &= \sfrac{1}{2}\!\left({\bf D}^j\!_m + \sfrac13\delta^j\!_m {\bf D}^2\right)\! f_{00} +
\partial_\eta\tilde{\mathsf{Q}}^{0j}\!_m,\\
\bar{\mathsf{R}}^{0j}\!_{km} &= 2{\bf D}_{[k}\tilde{\mathsf{Q}}^{0j}\!_{m]} ,\\
\bar{\mathsf{R}}^{ij}\!_{0m} &=  - 2{\bf D}^{[i}
\tilde{\mathsf{Q}}^{|0|j]}\!_{m} + 2Kf_0\!^{[i}\delta_m\!^{j]}, \\
\bar{\mathsf{R}}^{ij}\!_{km} &= -2\!\left({\bf D}_{[k}{\bf
D}^{[i} + K\delta_{[k}\!^{[i}\right)f_{m]}\!^{j]},
\end{align}
\end{subequations}
\begin{subequations}\label{Q_sf}
\begin{xalignat}{2}
\tilde{\mathsf{Q}}^0\!_{00} &= -\sfrac{1}{2}\partial_\eta
f_{00}, &\qquad \tilde{\mathsf{Q}}^0\!_{0 i} &=
-\sfrac{1}{2}{\bf D}_i f_{00},\\
\tilde{\mathsf{Q}}^0\!_{ij} &= \sfrac12
\partial_\eta f_{ij} -{\bf D}_{(i} f_{j)0},     &\qquad \tilde{\mathsf{Q}}^i\!_{0 0} &=
\partial_\eta f_0\!^i
- \sfrac12{\bf D}^{i} f_{00},\\
\tilde{\mathsf{Q}}_{i 0 j} &= \sfrac{1}{2}\partial_\eta f_{ij} - {\bf D}_{[i} f_{j]0},
&\qquad
\tilde{\mathsf{Q}}_{ijk} &= {\bf D}_{(j} f_{k)i} -
\sfrac{1}{2}{\bf D}_i f_{jk}.
\end{xalignat}
\end{subequations}
In summary, equations~\eqref{Einstein_source1}, in conjunction
with~\eqref{Einstein_source2}-\eqref{Q_sf}, give the general
expressions for the Einstein source terms.

\subsection{Gauge invariance and the Replacement Principle}\label{app:repl}

\subsubsection*{Gauge fields and gauge invariants}

In cosmological perturbation theory a second order gauge transformation can
be represented in coordinates as follows:
\be {\tilde x}^a = x^a +\epsilon {}^{(1)}\xi^a+
\sfrac12 \epsilon^2\left({}^{(2)}\xi^a+ {}^{(1)}\xi^a\,_{,b}{}^{(1)}\xi^b\right),   \ee
where ${}^{(1)}\!\xi^a$ and ${}^{(2)}\!\xi^a$ are independent dimensionless
background vector fields. Such a transformation induces a change in the first
and second order perturbations of a tensor field $A$ according to
\begin{subequations}\label{delta_A}
\begin{align}
{}^{(1)}\!\tilde{A}[\xi] &= {}^{(1)}\!A +
\pounds_{{}^{(1)}\!\xi}{}^{(0)}\!A,  \label{delta_A1}\\
{}^{(2)}\!\tilde{A}[\xi] &= {}^{(2)}\!A +
\pounds_{{}^{(2)}\!\xi}{}^{(0)}\!A +
\pounds_{{}^{(1)}\!\xi}\left(2{}^{(1)}\!A
+ \pounds_{{}^{(1)}\!\xi}{}^{(0)}\!A\right) ,
\label{delta_A2}
\end{align}
\end{subequations}
where $\pounds$ is the Lie derivative (see, e.g., Bruni {\it et
al} (1997), equations (1.1)--(1.3)).

One can impose restrictions on the tensor perturbations
${}^{(r)}\!\tilde{A}[\xi]$ by letting ${}^{(r)}\!\xi^a$ depend suitably on
the dimensionless perturbations $a^n\,{}^{(r)}\!{A}$ ($r=1,2$), a procedure
that can be referred to as perturbative gauge fixing. If the vector fields
${}^{(r)}\!\xi^a$ up to order $r$ are fully determined via the assumed
perturbative restrictions, we say that the gauge is fully fixed to order $r$.
An important special case of gauge fixing is \emph{order by order gauge
fixing}, in which the same conditions are imposed on the first and second (or
higher) order perturbations. When the gauge has been fully fixed to order
$r$, {\it all remaining perturbative quantities to order $r$ are rendered
gauge-invariant.}

The key features of our dimensionless version of Nakamura's method for
constructing gauge invariants up to second order are as follows (see Nakamura
(2007), and UW1, equations (5)--(8), for the linear case and further
discussion and references). Given a family of tensor fields $A(\epsilon)$
with $a^nA(\epsilon)$ dimensionless we define\footnote {Compare with
equations (2.26)--(2.27) in Nakamura (2007) and (2.34)--(2.35) in Nakamura
(2010). The factor of $a^n$ in our equations ensures that our expressions
${}^{(r)}\!{\bf A}[X],\, r=1,2,$ are dimensionless.}
\begin{subequations}\label{bold_A}
\begin{align} {}^{(1)}\!{\bf A}[X] &:= a^n \left({}^{(1)}\!A
- \pounds_{{}^{(1)}\!X} {}^{(0)}\!A \right), \label{bold_A1}\\
{}^{(2)}\!{\bf A}[X] &:= a^n \left({}^{(2)}\!A
- \pounds_{{}^{(2)}\!X}{}^{(0)}\!A
- \pounds_{{}^{(1)}\!X}\left(2A {}^{(1)} -
\pounds_{{}^{(1)}\!X}{}^{(0)}\!A\right)\right).\label{bold_A2}
\end{align}
\end{subequations}
Comparing~\eqref{bold_A} and~\eqref{delta_A} multiplied by $a^n$, reveals
that the equations have precisely the same form if we identify
$-{}^{(r)}\!X^a$ with ${}^{(r)}\!\xi^a$. Hence given a fully specified gauge
choice ${}^{(r)}\!\xi^a_\bullet$, if we choose ${}^{(r)}\!X^a_\bullet =
-{}^{(r)}\!\xi^a_\bullet$ then the ${}^{(r)}\!{\bf A}[X_\bullet]$ will be
gauge invariants that coincide with the
$a^n\,{}^{(r)}\!\tilde{A}[\xi_\bullet]$. In other words, imposing conditions
on the ${}^{(r)}\!{\bf A}[X]$ that fully determine the ${}^{(r)}\!X^a$
corresponds precisely to full gauge fixing. Due to the close relation between
the roles of ${}^{(r)}\!X^a$ and ${}^{(r)}\!\xi^a$ we refer to
${}^{(r)}\!X^a$ as \emph{gauge fields}, and we say that ${{}^{(1)}\!\bf
A}[X]$ and ${{}^{(2)}\!\bf A}[X]$ are the first and second order
\emph{dimensionless gauge invariants associated with ${}^{(1)}\! A$ and
${}^{(2)}\! A$, respectively, by $X$-compensation}.

We note in passing that another way of ensuring that the ${}^{(r)}\!{\bf
A}[X]$ are gauge-invariant expressions is to choose dimensionless fields
${}^{(r)}\!X^a$ that satisfy\footnote {These conditions arise in Nakamura's
work. See for example Nakamura (2007), equations (2.23) and (2.25).}
\begin{subequations}\label{delta X}
\begin{align}
\Delta {}^{(1)}\!X^a &= {}^{(1)}\!\tilde{X}^a - {}^{(1)}\!X^a = {}^{(1)}\!\xi^a, \label{delta_X1}\\
\Delta {}^{(2)}\!X^a &= {}^{(2)}\!\tilde{X}^a - {}^{(2)}\!X^a = {}^{(2)}\!\xi^a +
[{}^{(1)}\!\xi,{}^{(1)}\!X]^a. \label{delta_X2}
\end{align}
\end{subequations}
These conditions are obtained by applying~\eqref{delta_A} to~\eqref{bold_A}
and demanding that $\Delta {{}^{(1)}\!\bf A}[X] = 0$ and $\Delta
{{}^{(2)}\!\bf A}[X] = 0$. In UW1 we made use of~\eqref{delta_X1} in
constructing gauge invariants (see section 2.2). At second order, however,
the length of the computation makes it impractical to verify~\eqref{delta_X2}
directly.

The one-to-one correspondence between full gauge fixing and the Nakamura
approach makes the choice between them a matter of aesthetics and personal
preference.\footnote {The relation between gauge fixing and Nakamura's method
has also been discussed by Christopherson {\it et al} (2011).} We choose to
work in the formally `dimensionless gauge-invariant ${}^{(r)}\!{\bf A}[X]$
picture', but everything we do has a `fully gauge fixed picture' analogue
(replace $X$ with $-\xi$ and ${}^{(r)}\!{\bf A}[X]$ with
$a^n\,{}^{(r)}\!\tilde{A}[\xi]$).

In Appendix~\ref{app:gauge_field} we apply the Nakamura approach to the
metric tensor to construct gauge fields and gauge invariants.

\subsubsection*{The Replacement Principle}

The equivalence of gauge fixing and the Nakamura approach is expressed in the
Replacement Principle. We present two versions, one for the Riemann and
Einstein tensors, and one for the stress-energy tensor.

The dependence of the perturbations of the Riemann tensor on the metric
perturbations can be written symbolically in the form~\eqref{rie_operator},
using the geometric operators:
\begin{subequations} \label{Rie_RP}
\be a^2{}^{(1)}\!R^{ab}\!_{cd} =  {\mathsf R}^{ab}\!_{cd}({}^{(1)}\!f),\qquad
a^2{}^{(2)}\!R^{ab}\!_{cd} = \mathsf{R}^{ab}\!_{cd}({}^{(2)}\!f) + {\cal
R}^{ab}\!_{cd}({}^{(1)}\!f,{}^{(1)}\!f), \ee
where ${}^{(r)}\!f$ is shorthand for ${}^{(r)}\!f_{ab}$, with $r=1,2$. The
Replacement Principle for the Riemann curvature states that the gauge
invariants associated with ${}^{(r)}\!R^{ab}\!_{cd}$ and with
${}^{(r)}\!f_{ab}$ by $X$-compensation are related by {\it the same}
operators:
\be {}^{(1)}\!{\bf R}^{ab}\!_{cd}[X] =  {\mathsf R}^{ab}\!_{cd}({}^{(1)}{\bf
f}),\qquad {}^{(2)}\!{\bf R}^{ab}\!_{cd}[X] =
\mathsf{R}^{ab}\!_{cd}({}^{(2)}\!{\bf f}) + {\cal R}^{ab}\!_{cd}({}^{(1)}{\bf
f},{}^{(1)}{\bf f}),
\ee
\end{subequations}
where ${}^{(r)}{\bf f}$ is shorthand for ${}^{(r)}{\bf f}_{ab}[X]$. A similar
result for the Einstein tensor can be derived from the above
using~\eqref{ein}.

This Replacement Principle has its origins in the work of Nakamura, although
he does not state it  explicitly in the above form. See, for example,
Nakamura (2010), equations (B9)--(B13), in conjunction with equations (2.36)
and (2.37).

The stress-energy tensor of a perfect fluid, as given by~\eqref{pf}, can be
viewed as a function of the variables $F=(\rho,p,u_a,g_{ab})$. The
perturbations of the stress-energy tensor can be written symbolically in the
form:
\begin{subequations} \label{T_RP}
\be \label{T_gv}  a^2{}^{(1)}\!{ T}^{a}\!_{b} =
\mathsf{T}^{a}\!_{b}(^{(1)}\!{F}),\quad a^2{}^{(2)}\!{T}^{a}\!_{b} =
\mathsf{T}^{a}\!_{b}(^{(2)}\!{F}) +
{\mathcal{T}}^{a}\!_{b}(^{(1)}\!{F},{}^{(1)}\!{F}),    \ee
where $ \mathsf{T}^{a}\!_{b}$ is the linear leading order operator and
${\mathcal{T}}^{a}\!_{b}$ is the quadratic source term operator, and
${}^{(r)}\!F=({}^{(r)}\!\rho,{}^{(r)}\!p,{}^{(r)}\!u_a,{}^{(r)}\!g_{ab})$,
with $r=1,2$. The Replacement Principle for the stress-energy tensor states
that the gauge invariants associated with ${}^{(r)}\!T^a\!_b$ and with
${}^{(r)}\!F$ by $X$-compensation are related by {\it the same} operators:
\be \label{T_gi}  {}^{(1)}\!{\bf T}^{a}\!_{b}[X] =
\mathsf{T}^{a}\!_{b}(^{(1)}{\bf F}),\qquad {}^{(2)}\!{\bf T}^{a}\!_{b}[X] =
\mathsf{T}^{a}\!_{b}({}^{(2)}\!{\bf F}) +
{\mathcal{T}}^{a}\!_{b}({}^{(1)}\!{\bf F},{}^{(1)}\!{\bf F} ),    \ee
\end{subequations}
where ${}^{(2)}\!{\bf F}$ is shorthand for ${}^{(2)}\!{\bf F}[X]$. This
result can be deduced from Nakamura (2007).\footnote {His equations
(4.97)--(4.102) correspond to our equations~\eqref{T_RP} although it requires
close scrutiny to conclude that his equations can be written in the operator
form of~\eqref{T_RP}.}

\subsection{Construction of the gauge fields $X^a$}\label{app:gauge_field}

We begin with the metric mode decomposition~\eqref{f_decomp}. This
decomposition assumes that the inverse operators (Green's functions) ${\bf
D}^{-2}$, $({\bf D}^2 + 2K)^{-1}$ and $({\bf D}^2 + 3K)^{-1}$ exist (see UW1,
and Nakamura (2007), page 19, for further discussion), as seen explicitly
when one extract the various modes. In the present context this is
accomplished by means of the following \emph{mode extraction
operators},\footnote {Note that  ${\bf D}^2$ and ${\bf D}_{ij}$ are defined
in equation~\eqref{2order_D}, and that $\la ij\ra$ is defined in
footnote~\ref{hat_angle}.}
\begin{subequations}\label{modeextractop}
\begin{xalignat}{2}
{\cal S}^{i} &= {\bf D}^{-2}{\bf D}^i ,&\quad {\cal S}^{ij} &=
\sfrac32 {\bf D}^{-2}\!\left({\bf D}^2 + 3K\right)^{-1}{\bf
D}^{ij},\\
{\cal V}_i\!^{j} &= \delta_i\!^j - {\bf D}_i{\cal S}^j, &\quad
{\cal V}_i\!^{jk} &=
\left({\bf D}^2 + 2K\right)^{-1} {\cal V}_i\!^{\la j}{\bf D}^{k\ra},\\
{\cal T}_{ij}\!^{km} &= \delta_i\!^{\la k}\delta_j\!^{m\ra} -
{\bf D}_{(i}{\cal V}_{j)}\!^{km} - {\bf D}_{ij}{\cal S}^{km}.
&&
\end{xalignat}
\end{subequations}
Applying these operators to~\eqref{f_decomp} gives
\begin{subequations}\label{f_decomp2}
\begin{xalignat}{3}
\varphi &= -\sfrac12 f_{00}, &\quad B &= {\cal S}^{i}f_{0i}
,&\quad B_i &= {\cal V}_i\!^{j} {f}_{0j},\\
C &= \sfrac12\,{\cal S}^{ij} {f}_{ij}, &\quad C_i &= {\cal
V}_i\!^{jk}{f}_{jk}, &\quad
C_{ij} &= \sfrac12\,{\cal T}_{ij}\!^{km}{f}_{km},\\
 \psi &= - \sfrac16(f - {\bf D}^2{\cal S}^{ij}{f}_{ij}) . &&
\end{xalignat}
\end{subequations}
Analogous relations hold for the modes of ${}^{(2)}\!f_{ab}$.
Explicitly extracting modes by means of the mode extraction
operators~\eqref{modeextractop} becomes essential at second
order, and these operators are therefore used frequently in
this paper.

In order to construct gauge invariants associated with the metric
perturbations we apply~\eqref{bold_A} with $n=-2$ to
$A_{ab}=g_{ab}=a^2 {\bar g}_{ab}$ and define\footnote
{Here
${}^{(0)}\!A_{ab}={}^{(0)}\!g_{ab}=a^2\gamma_{ab}$,
${}^{(r)}\!A_{ab}={}^{(r)}\!g_{ab} = a^2\, {}^{(r)}\!f_{ab},$
and we denote the dimensionless gauge invariant associated with
${}^{(r)}{g}_{ab}$ by ${}^{(r)}{\bf f}_{ab}[X],\, r=1,2$.}
\begin{subequations}\label{bold_f}
\begin{align}
{\bf f}_{ab}[X] &:= f_{ab} - a^{-2}\pounds_{{}^{(1)}\!X}\!\left(a^2\gamma_{ab}\right), \label{bold_f1}\\
{}^{(2)}\!{\bf f}_{ab}[X] &:= {}^{(2)}\!f_{ab} - a^{-2}\pounds_{{}^{(2)}\!X}\!\left(a^2\gamma_{ab}\right) +
{\mathcal F}_{ab}[X],\label{bold_f2}
\end{align}
where
\be {\mathcal F}_{ab}[X] := - a^{-2}\pounds_{{}^{(1)}\!X}\!\left(2a^2 f_{ab} -
\pounds_{{}^{(1)}\!X}\!\left(a^2\gamma_{ab}\right)\right)\!. \label{F2X2} \ee
\end{subequations}
We perform a mode decomposition of ${\bf f}_{ab}[X]$ and ${}^{(2)}{\bf
f}_{ab}[X]$ using equations~\eqref{f_decomp} as a model, and introduce an
obvious notation.\footnote{For example,
${}^{(2)}\varphi\rightarrow{}^{(2)}\Phi[X], {}^{(2)}B\rightarrow{}^{(2)}{\bf
B}[X]$, as in equations~\eqref{modefX}.  } This mode decomposition enables
one to determine the gauge fields uniquely, as follows.

We begin by expressing~\eqref{bold_f1} and~\eqref{bold_f2} in terms of the
modes as defined by~\eqref{f_decomp}, and the mode decomposed spatial vectors
${}^{(r)}\!X_i = {\bf D}_i{}^{(r)}\!X + {}^{(r)}\!\tilde{X}_i$,
which gives
\begin{subequations}\label{modefX}
\begin{align}
{}^{(r)}\!\Phi[X] &= {}^{(r)}\!\varphi -
(\partial_\eta + {\cal H}){}^{(r)}\!X^0 -\sfrac12
{\mathcal F}_{00}[X], \label{PhiX}\\
{}^{(r)}\!{\bf B}[X] &= {}^{(r)}\!B + {}^{(r)}\!X^0 - \partial_\eta{}^{(r)}\!X + {\cal S}^{i}{\mathcal F}_{0i}[X], \label{BX}\\
{}^{(r)}\!{\bf B}_i[X] &= {}^{(r)}\!B_i - \partial_\eta{}^{(r)}\!\tilde{X}_i + {\cal V}_i\!^{j}{\mathcal F}_{0j}[X], \label{BiX}\\
{}^{(r)}\!{\bf C}[X] &= {}^{(r)}\!C - {}^{(r)}\!X + \sfrac12\,{\cal S}^{ij} \hat{{\mathcal F}}_{ij}[X], \label{CX}\\
{}^{(r)}\!{\bf C}_i[X] &= {}^{(r)}\!{C}_i - {}^{(r)}\!\tilde{X}_i + {\cal V}_i\!^{jk}\hat{{\mathcal F}}_{jk}[X], \label{CiX}\\
{}^{(r)}\!{\bf C}_{ij}[X] &= {}^{(r)}\!{C}_{ij} +
\sfrac12\,{\cal T}_{ij}\!^{km}\hat{{\mathcal F}}_{km}[X], \label{CijX}\\
{}^{(r)}\!\Psi[X] &= {}^{(r)}\!\psi + {\cal H}{}^{(r)}\!X^0 -
\sfrac16({\mathcal F}^k\!_k[X] - {\bf D}^2{\cal S}^{ij}\hat{{\mathcal F}}_{ij}[X]), \label{PsiX}
\end{align}
\end{subequations}
where $r=1,2$, and the source terms ${\mathcal F}_{ab}[X]$ do not appear when $r=1$.

We next determine ${}^{(r)}\!X$ and ${}^{(r)}\!\tilde{X}_i$ uniquely by
imposing the conditions ${}^{(r)}\!{\bf C}[X]=0$ and ${}^{(r)}\!{\bf
C}_i[X]=0$ in~\eqref{CX} and~\eqref{CiX}, respectively, which lead to
\begin{subequations}\label{XC}
\begin{xalignat}{2}
{}^{(1)}\!X &= {}^{(1)}\!C, &\,
{}^{(1)}\!\tilde{X}_i &= {}^{(1)}\!{C}_i, \label{XC1}\\
{}^{(2)}\!X &= {}^{(2)}\!C  + \sfrac12\,{\cal S}^{ij}
\hat{{\mathcal F}}_{ij}[X], &\, {}^{(2)}\!\tilde{X}_i &=
{}^{(2)}\!{C}_i + {\cal
V}_i\!^{jk}\hat{{\mathcal F}}_{jk}[X].\label{XC2}
\end{xalignat}
\end{subequations}
At this stage the mode decomposition for ${}^{(r)}{\bf f}_{ab}[X], r=1,2$
assumes the form\footnote
{This choice of the spatial gauge is essentially that made by Noh and Hwang (2004).
See their equation (259) and the discussion on page 37.}
\begin{subequations} \label{bold_f_split}
\begin{align}
{}^{(r)}{\bf f}_{00}[X] &= -2  {}^{(r)}\Phi[X] \, ,\\
{}^{(r)}{\bf f}_{0 i}[X] &= {\bf D}_i {}^{(r)}{\bf B}[X] + {}^{(r)}{\bf B}_i[X]\, ,\\
{}^{(r)}{\bf f}_{ij}[X] &= -2{}^{(r)}\Psi[X] \gamma_{ij} + 2{}^{(r)}{\bf C}_{ij}[X]\, .
\end{align}
\end{subequations}
It remains to determine ${}^{(r)}\!X^0$, where ${}^{(1)}\!X^0$ together with
${}^{(1)}\!X = {}^{(1)}\!C$ and ${}^{(1)}\!\tilde{X}_i = {}^{(1)}\!{C}_i$
form ${}^{(1)}\!X^a$ which is to be inserted in $\hat{{\mathcal F}}_{ij}[X]$
in~\eqref{XC2}. There are two ways to determine ${}^{(r)}\!X^0$ algebraically
in a unique manner. The first way is to set ${}^{(r)}\!{\bf B}[X]=0$, which
via~\eqref{BX} with~\eqref{XC} inserted yields
\begin{subequations}\label{PX}
\begin{align}
{}^{(1)}\!X^0 &= {}^{(1)}\!X_\mathrm{p}^0 := -{}^{(1)}\!B  +
\partial_\eta{}^{(1)}\!C,\label{PX1}\\
{}^{(2)}\!X^0 &= {}^{(2)}\!X_\mathrm{p}^0 := -{}^{(2)}\!B  +
\partial_\eta{}^{(2)}\!C - {\cal S}^{i}{\mathcal F}_{0i}[X_\mathrm{p}].\label{PX2}
\end{align}
\end{subequations}
Note that ${}^{(1)}\!X_\mathrm{p}^0$ together with~\eqref{XC1} yields
${}^{(1)}\!X^a_\mathrm{p}$. Substituting ${}^{(1)}\!X^a_\mathrm{p}$ into
$\hat{{\mathcal F}}_{ij}[X]$ and ${\mathcal F}_{0i}[X_\mathrm{p}]$
in~\eqref{XC2} and~\eqref{PX2}, respectively, gives
${}^{(2)}\!X^a_\mathrm{p}$. We refer to the gauge fields
${}^{(r)}\!X_\mathrm{p}^a$ as the \emph{Poisson gauge fields} since they
result in the \emph{Poisson gauge invariants}:
\begin{equation} \label{Poisson}
{}^{(r)}\!\Phi := {}^{(r)}\!\Phi[X_\mathrm{p}],\qquad
{}^{(r)}\!\Psi := {}^{(r)}\!\Psi[X_\mathrm{p}],\qquad
{}^{(r)}\!{\bf B}_i[X_\mathrm{p}],\qquad {}^{(r)}\!{\bf
C}_{ij}[X_\mathrm{p}] ,
\end{equation}
when~\eqref{XC} and~\eqref{PX} are inserted into~\eqref{modefX}.

The second way is to set ${}^{(r)}\!\Psi[X]=0$, which via~\eqref{PsiX}
with~\eqref{XC1} inserted gives
\begin{subequations}\label{CuX}
\begin{align}
{}^{(1)}\!X^0 &= {}^{(1)}\!X_\mathrm{c}^0 := - {\cal
H}^{-1}{}^{(1)}\!\psi,\label{CX1}\\
{}^{(2)}\!X^0 &= {}^{(2)}\!X_\mathrm{c}^0 := - {\cal
H}^{-1}{}^{(2)}\!\psi +
\sfrac16({\mathcal F}^k\!_k[X_\mathrm{c}] - {\bf D}^2{\cal
S}^{ij}\hat{{\mathcal F}}_{ij}[X_\mathrm{c}]),\label{CX2}
\end{align}
\end{subequations}
where ${}^{(1)}\!X_\mathrm{c}^0$ together with~\eqref{XC1} yields
${}^{(1)}\!X^a_\mathrm{c}$. Inserting ${}^{(1)}\!X^a_\mathrm{c}$ into
$\hat{{\mathcal F}}_{ij}[X]$ in~\eqref{XC2} and~\eqref{CX2}, and into
${\mathcal F}^k\!_k[X_\mathrm{c}]$ in~\eqref{CX2}, gives
${}^{(2)}\!X^a_\mathrm{c}$. We refer to the gauge fields
${}^{(r)}\!X_\mathrm{c}^a$ as the \emph{uniform curvature gauge fields} since
they result in the \emph{uniform curvature gauge invariants}\footnote{We note
in passing that that Christopherson {\it et al} (2011) have given the
relation between the uniform curvature metric gauge invariants $^{(2)}\!{\bf
A}, ^{(2)}\!{\bf B}$ and the Poisson gauge invariants $^{(2)}\!{\Phi},
^{(2)}\!{\Psi}$ (see their equations (4.56) and (4.58)).}
\begin{equation}
{}^{(r)}\!{\bf A} := {}^{(r)}\!\Phi[X_\mathrm{c}],\qquad
{}^{(r)}\!{\bf B} := {}^{(r)}\!{\bf B}[X_\mathrm{c}],\qquad
{}^{(r)}\!{\bf B}_i[X_\mathrm{c}],\qquad {}^{(r)}\!{\bf
C}_{ij}[X_\mathrm{c}] ,
\end{equation}
when~\eqref{XC} and~\eqref{CuX} are inserted into~\eqref{modefX}.

Note that at first order ${}^{(1)}\!{\bf C}_{ij}[X] = {}^{(1)}\!{\bf
C}_{ij}$, i.e., ${}^{(1)}\!{\bf C}_{ij}$ is independent of ${}^{(1)}\!X^a$,
while at second order ${}^{(2)}\!{\bf C}_{ij}[X]$ depends on $X$, so that
${}^{(2)}\!{\bf C}_{ij}[X_\mathrm{p}]$ and ${}^{(2)}\!{\bf
C}_{ij}[X_\mathrm{c}]$ are unequal.

\section{Source terms in the uniform curvature gauge}\label{app:ucg}

 We make the replacement
$f_{ab} \rightarrow {\bf f}_{ab}[X_{\mathrm c}]$ in the
expressions~\eqref{Einstein_source1} for the Einstein source terms using the
metric perturbation~\eqref{bf_curv} in the uniform curvature gauge
subject to the restriction~\eqref{m_assump1}, {\it i.e.}
\be {\bf f}_{00}[ X_{\mathrm{c}}] = -2{\bf A} ,\qquad
{\bf f}_{0i}[ X_{\mathrm{c}}] = {\bf D}_i {\bf B}, \qquad {\bf
f}_{ij}[ X_{\mathrm{c}}] = 0. \ee
The constituent terms, as
given by equations~\eqref{Einstein_source2}--\eqref{Q_sf}, can be evaluated
separately. This calculation yields
the following expressions for the source terms in the uniform curvature gauge,
assuming a purely scalar linear perturbation:

\begin{subequations}
\begin{align}
\begin{split}
\hat{\mathcal{G}}_{ij}({\bf f}_{\mathrm{c}},{\bf f}_{\mathrm{c}}) &= 2\!\left[2{\bf A}{\bf D}_{ij}({\bf A} + {\mathcal L}_B{\bf B}) +
\left(\partial_\eta {\bf A} + \sfrac13{\bf D}^2{\bf B}\right){\bf D}_{ij}{\bf B}\right.\\
& \left. +\, ({\bf D}_{\la i}{\bf A}){\bf D}_{j\ra}({\bf A} + 2{\cal H}{\bf B})
 - ({\bf D}_{k\la i}{\bf B})\,{\bf D}^k\!_{j\ra}{\bf B} - K({\bf D}_{\la i}{\bf B})\,{\bf D}_{j\ra}{\bf B}  \right],\end{split} \\
\begin{split}{\mathcal{G}}({\bf f}_{\mathrm{c}},{\bf f}_{\mathrm{c}}) &=
-\sfrac13(1 + 3{\cal C}_G^2){\mathcal R} -\, \sfrac13{\mathcal{W}}^k\!_k
+ 2{\cal H}{\cal L}_A(-4{\bf A}^2 + ({\bf D}{\bf B})^2 )  \\
& \quad -\, \sfrac43\!\left[2{\bf A}{\bf D}^2({\bf A} + {\mathcal L}_B{\bf B}) + (\partial_\eta{\bf A}-4{\cal H}{\bf A})\,{\bf D}^2{\bf B}
+ ({\bf D}{\bf A})^2\right],
\end{split}\\
{\mathcal{G}}_{i}({\bf f}_{\mathrm{c}},{\bf f}_{\mathrm{c}}) &= {\bf D}_i{\mathcal R}-
3{\cal H}\bar{\mathcal{G}}^{0}\!_{i},\\
{\mathcal{G}}^0\!_{i}({\bf f}_{\mathrm{c}},{\bf f}_{\mathrm{c}}) &=
 - 2{\cal H}{\bf D}_i(-4{\bf A}^2 + ({\bf D}{\bf B})^2) + \bar{\mathcal{G}}^{0}\!_{i},
\end{align}
where
\begin{align}
{\mathcal{W}}^k\!_k &=
4{\cal H}\!\left[2{\bf A}\,{\bf D}^2{\bf B} + ({\bf D}^k{\bf A})\,{\bf D}_k{\bf B}\right],\\
{\mathcal R} &={\mathcal W}^k\!_k
-({\bf D}^k\!_m {\bf B})\,{\bf D}^m\!_k{\bf B} + \sfrac23({\bf D}^2{\bf B})^2 - 4K({\bf D}{\bf B})^2,\\
\bar{\mathcal{G}}^{0}\!_{i} &=
- 2 ({\bf D}_j{\bf A})\,{\bf D}^j\!_i {\bf B} +
\sfrac43({\bf D}_i{\bf A})\,{\bf D}^2{\bf B} - 8K{\bf A}\,{\bf D}_i{\bf B}.
\end{align}
\end{subequations}
We note that the source terms in the Poisson gauge as given by~\eqref{special_source} were derived by imposing the
restriction $\Phi=\Psi$, which led to significant simplification. If we impose the corresponding
restriction in the uniform curvature case, namely ${\bf A} =- {\mathcal L}_B{\bf B}$
(see equations (38a) and (40a) in UW2), then the
above expressions simplify somewhat, as can be seen by inspection.

\end{appendix}

\section*{References}

\noindent Acquaviva, V., Bartolo, N., Matarrese, S. and Riotto, A. (2003)	
Gauge-invariant second-order perturbations and non-Gaussianity from inflation,
{\it Nuclear Physics B} {\bf 667}, 119-148.\\

\noindent Ananda, K.N, Clarkson, C.  and Wands, D.  (2007)
The cosmological gravitational wave background from primordial density
perturbations, {\it Phys. Rev. D} {\bf 75}, 123518.\\

\noindent Bardeen, J. M. (1980) Gauge-invariant cosmological
perturbations, {\it Phys. Rev. D} {\bf 22}, 1882-1905.\\

\noindent Bartolo, N., Matarrese, S., Pantano, O. and Riotto, A. (2010a)	
Second-order matter perturbations in a $\Lambda$CDM cosmology and non-Gaussianity,
 {\it Class. Quant. Grav.} {\bf 27}, 124009. \\

\noindent Bartolo, N., Matarrese, S. and Riotto, A. (2010b)	
Non-Gaussianity and the Cosmic Microwave Background Anisotropies,
{\it Advances in Astronomy}, {\bf 2010}, 157079, arXiv:1001.3957 [astro-ph.CO]. \\

 \noindent	Bartolo, N., Komatsu, E., Matarrese, S. and Riotto, A. (2004a)
 Non-Gaussianity from inflation: theory and observations,
 {\it Physics Reports} {\bf 402}, 103-266. \\

 \noindent	Bartolo, N., Matarrese, S. and Riotto, A. (2004b)
Enhancement of Non-Gaussianity after Inflation,
{\it JEHP} {\bf 06404}, 006. \\

\noindent Baumann, D., Steinhardt, P. and Takahashi, K. (2007)
Gravitational wave spectrum induced by primordial scalar
perturbations, {\it Phys. Rev. D} {\bf 76}, 084019.\\

\noindent Bruni, M., Dunsby, P.K.S. and Ellis, G.F.R. (1992)
Cosmological perturbations and the meaning of gauge-invariant
variables, {\it Astrophysical J.} {\bf 395}, 34-53.\\

\noindent Bruni, M., Matarrese, S., Mollerach, S. and Sonego,
S. (1997) Perturbations of spacetime: gauge transformations and
gauge-invariance at second order and beyond,
{\it Class. Quant. Grav.} {\bf 14}, 2585-2606.\\

\noindent Christopherson, A. J., Malik, K. A., Matravers, D. R.
and Nakamura, K. (2011) Comparing different formulations of
nonlinear perturbation theory,
 {\it Class. Quant. Grav.} {\bf 28}, 225024. \\

\noindent Christopherson, A. J. and Malik, K. A. (2009)
Practical tools for third order cosmological perturbations,
{\it JCAP} {\bf 0911}, 012.  \\

\noindent Hwang, J-C., Noh, H. and Gong, J-O. (2012), Second order solutions of cosmological
perturbations in the matter dominated era, arXiv:1204.3345. \\

\noindent Hwang, J. and Noh, H (2007)
Second-order perturbations of cosmological fluids: relativistic effects of pressure,
multicomponent, curvature, and rotation
{\it Phys. Rev. D} {\bf 76}, 103527.  \\

\noindent Huston, I. and Malik, K.A. (2011), Second order
perturbations during inflation beyond slow roll,
 {\it JCAP} {\bf 1110}, 029. \\

\noindent Kodama, H. and Sasaki, M. (1984) Cosmological
Perturbation Theory, {\it Prog. Theoret. Phys. Suppl. } {\bf 78}, 1-166.\\

\noindent  Malik, K. A. (2007) A not so short note on the
Klein-Gordon
equation at second order, {\it JCAP} {\bf 0703}, 004.  \\

\noindent Malik, K. A. and Wands, D. (2004)
Evolution of second-order cosmological perturbations,
{\it Class. Quant. Grav.} {\bf 21L}, 65-70.\\

\noindent Malik, K. A., Seery, D. and Ananda, K. N. (2008) Different approaches
to the second order Klein-Gordon equation,  {\it Class. Quant. Grav.} {\bf 25}, 175008.\\

\noindent Mukhanov, V. F., Feldman, H. A. and Brandenberger,
R. H. (1992) Theory of cosmological perturbations, {\it Physics Reports}
{\bf 215}, 203-333.\\

\noindent Nakamura, K. (2003) Gauge Invariant Variables in
Two-Parameter Nonlinear Perturbations, {\it Prog. Theor. Phys.}
{\bf 110}, 723-755. \\

\noindent Nakamura, K. (2006) Gauge-invariant Formulation  of
the Second-order Cosmological  Perturbations, {\it Phys. Rev.
D} {\bf 74}, 101301.\\

\noindent Nakamura, K. (2007) Second Order Gauge Invariant
Cosmological Perturbation Theory, {\it Prog. Theor. Phys.} {\bf
117}, 17-74. \\

\noindent Nakamura, K. (2010) Second-order Gauge-invariant
Cosmological Perturbation Theory: Current Status, {\it Advances
in Astronomy} {\bf 2010}, 576273. \\

\noindent Noh, H. and Hwang, J. (2004) Second order
perturbations of the Friedmann world model, {\it Phys. Rev. D}
{\bf 69} 104011. \\

\noindent  Pitrou, C., Uzan, J-P. and Bernardeau, F. (2010)
The cosmic microwave background bispectrum from
the non-linear evolution of the cosmological
perturbations,  {\it JCAP} {\bf1007}, 003.  \\

\noindent Uggla, C and Wainwright, J. (2011) Cosmological
Perturbation theory revisited, {\it Class. Quant. Grav.} {\bf 28},
175017 .\\

\noindent Uggla, C and Wainwright, J. (2012) Dynamics of
cosmological scalar perturbations, {\it Class. Quant. Grav.} {\bf 29},
105002 . \\

\noindent Wainwright, J. and Ellis, G.F.R. (1997) {\it
Dynamical systems in cosmology}, Cambridge University Press. \\

\end{document}